%% file: main.tex
\documentclass[preprint, 12pt]{elsarticle}

\input{preamble}
\begin{document}
\input{frontmatter}
\section{Introduction}
\label{Introduction}
The spray atomization process involves complex multi-phase breakup phenomena which are highly pertinent in many industrial applications, such as fuel injection, gas turbines, spray painting, fire protection, aero-engines and propulsion systems. Extensive research has been conducted on such industrial sprays that function at high pressure \citep{treleaven2022coupling, rostami2021velocity}. Using instability theory, numerous mathematical models have been developed to depict spray characteristics, eliminating the requirement of explicitly modelling the atomization process \citep{hossainpour2009investigation, yu2016modeling, schmidt1999pressure}. However, only a handful of studies investigate atomization from a nasal spray limited to discrete phase models \citep{Inthavong2008,Inthavong2011,Fung2013,Kolanjiyil2021,Calmet2019}, where the operating pressure is much lower compared with industrial sprays. 

The spray generated by nasal spray devices is typically formed from a simplex atomizer (also called pressure swirl atomizer), characterized by tangential ports. These ports direct the liquid into a swirl chamber tangentially to 
promote strong rotational motion producing an air core region. As a result, the liquid exits the nozzle of the atomizer as a conical liquid sheet \citep{Inthavong2012} under the action of both axial and tangential forces \citep{zheng2023numerical}. This provides good atomization performance, due to the increased air-liquid interactions \citep{Rizk1987} produced from the hollow conical liquid sheet discharged from the nozzle \citep{ Taylor1948, Moon2007,Dombrowski1969,Rizk1987,Rizk1985}.

The liquid sheet once formed is subjected to aerodynamic instabilities that cause it to disintegrate into ligaments and eventually break up into droplets. The primary function of a nasal spray is to transform the drug-infused liquid into finely atomized droplets, thus increasing the ratio of surface area to volume to provide better bioactivity for the treatment of nasal diseases and transport the droplets into the nasal cavity. As in other applications, the spray characteristics are affected by the air core developed due to the low pressure in the axial direction of the nozzle \citep{Liu2017}. The most commonly investigated characteristics are droplet size distribution, velocity magnitude, mass distribution in the spray cone, spray cone angle and interaction of droplets with the ambient air \citep{ Cheng2001,Suman2002,Dayal2004,Guo2006,Foo2007,Guo2008,Liu2010,Van2022}. 

The drug delivery efficacy of a nasal spray device depends on many factors, such as device design, device handling and the physical properties of the drug formulation \citep{Suman2002,Dayal2004}. Inertial impaction causes 90\% of the droplets to deposit in the anterior region of the nasal cavity thus limiting the aerosol that gets delivered to the targeted nasal turbinates \citep{Newman1988,Kimbell2007,Kelly2004}. Improving the delivery efficacy, therefore, requires a good understanding of primary atomization and secondary break-up of a nasal spray. With a nasal spray, the drug droplets need to be delivered under ambient conditions, unlike fuel injection where the fuel is introduced under high pressure and controlled chamber pressure which assists atomization. Due to this limitation, the nasal spray operates in the low Weber number and low gas-liquid momentum ratio regime. This regime of break-up is defined as axisymmetric break-up since the break-up is dominated by the liquid momentum \citep{Rajamanickam2017}. Nasal spray–wall interaction and post-deposition liquid motion are other essential aspects that impact drug bioactivity. CFD investigations have unveiled notable impacts of the spray momentum on both the movement of liquid on the surface and the motion of the surface film caused by the airflow resulting from shear stress and gravity \citep{Kolanjiyil2022,Si2021}. Shang et al. developed a novel approach combining Computational Fluid Dynamics (CFD) techniques with a 1-D mucus diffusion model to better predict nasal spray drug absorption \citep{Shang2021}.

Furthermore, the effective administration of the nasal spray is constrained by the patients' device handling ability and the actuation force \citep{Gao2020,Tong2016,Basu2020}. Actuation force is another limiting factor in a nasal spray application unlike the applied injection pressure used in fuel spray application to enhance atomization. Therefore, device design is an essential consideration to be accounted for in the nasal drug delivery application to improve atomization. 

Several studies have been carried out using high-speed imaging to gain insight into the near-field characteristics of a nasal spray. Inthavong et al. used high-speed photography and a particle droplet image analyzer to provide greater insight into the physical mechanisms of nasal spray atomization \citep{Inthavong2014}. It was concluded that increasing actuation pressure produced more rapid atomization with increased spray velocity, decreased droplet size and decreased spray time. Fung et al. identified three main phases of spray development (pre-stable, stable, and post-stable) that can be correlated by visualizing the spray width \citep{Fung2013}. Liu et al. evaluated droplet velocity and size from nasal spray devices using phase Doppler anemometry (PDA) \citep{Liu2010}.  For nasal spray application,  Shrestha et al. adapted image processing techniques to extract several other spray parameters, such as break-up length, dispersion angle, break-up radius and cone angle, which contribute to the discrete phase modelling (secondary break-up) numerical inputs for nasal spray application \citep{Shrestha2020}. This study was limited to a single over-the-counter nasal spray device. Later, \cite{Van2022} further extended this study with four different types of commercially-available nasal sprays used in an ENT (Ear Nose Throat) clinic and provided extensive data on nasal spray characteristics \citep{Van2022}. 

To date, no numerical studies can be found for the nasal spray application that resolves all the complex physics involved in primary atomization. Although, there are studies that have investigated breakup mechanisms from swirl atomizers \citep{patil2021air, sahu2022formation, kuo2022simulation} and other atomizers such as hollow cones \citep{di2022computational}, air-blast \citep{patil2021air}, and co-axial \citep{kumar2020liquid, charalampous2019proper}. Further computational studies of nasal spray applications can be used to provide a deeper understanding of the primary breakup of the liquid sheet which influences and secondary break-up of a nasal spray and eventually the droplet characteristics.

In this study, a high-fidelity numerical simulation was performed to characterize the liquid sheet formation and its disintegration into droplets for a low-pressure application of a nasal spray. The study aims to provide insight into the internal and external near-nozzle spray characterization of a continuous spray and to establish good validation against the experimental data. In addition, this study provides a detailed understanding of the mechanism of the liquid sheet, ligament and droplet formation in a swirling liquid jet of a nasal spray by using a novel VOF-DPM transition method supplemented by an AMR (adaptive mesh refinement) technique. The outcome of the study serves as a benchmark to idealize the parameters for the development of a spray atomization model for nasal application. 

\section{Method}
\subsection{Numerical methods and framework}
\subsubsection{Volume of Fluid (VOF) modelling}
The Volume of Fluid (VOF) free-surface modelling technique was used to track the gas-liquid interface \citep{Hirt1981}. The conservation equations for mass and momentum are:

\begin{equation}\label{NS-equation}
\frac{\partial\rho}{\partial{t}}+\nabla.(\rho\vec{u})=0
\end{equation}

\begin{equation}
\frac{\partial}{\partial{t}}\left(\rho\vec{v}\right)+\nabla.\left(\rho\vec{v}\otimes\vec{v}\right)=-\nabla{p}+\nabla.\left[\mu\left(\nabla\vec{v}+\nabla{\vec{v}}^T\right)\right]+\rho\vec{g}+\vec{F}
\end{equation}

where $\vec{v}$ is the velocity vector, $t$ is the time, $p$ is the static pressure, $\rho$ is the fluid density, $\mu$ is the dynamic viscosity, $\vec{g}$ is the gravity vector and $\vec{F}$ is an additional force which represents surface tension. The volume fraction $\alpha_l$ is used to track the movement of the gas-liquid interface which has a value of one in the liquid and zero in the gas phase. It is calculated via
\begin{equation}\label{interface}
\frac{\partial\alpha_l}{\partial{t}}+\vec{u}.\nabla\alpha_l=0
\end{equation}
In the transport equations, the material properties of the two phases are determined by the presence of the component phases in each control volume and are represented by the subscripts $l$ and $g$ so that, for example, the density in each cell is given by
\begin{equation}
\rho=\alpha_l\rho_l+(1-\alpha_l)\rho_g
\end{equation}
The surface tension force ($\vec{F}$) is modelled using the continuum surface force approach \citep{Brackbill1992}, with the surface curvature computed from local gradients of the surface normal at the interface. The model equations are

\begin{equation}
\vec{F}=\sigma\kappa\nabla\alpha_l
\end{equation}
\begin{equation}
\kappa=-\nabla.\left(\frac{\nabla\alpha_l}{\left|\nabla\alpha_l\right|}\right)
\end{equation}
where $\kappa$ is the surface curvature.

\subsubsection{Turbulence modelling}
The Stress-Blended Eddy Simulation (SBES) model was used to capture the local turbulence in the region of the jet. Hybrid RANS-LES models, such as SBES, allow a RANS model to be used in the near wall region while Large Eddy Simulation (LES) is used for the flow away from the wall. The subgrid-scale (SGS) stresses resulting from the filtering operation are modelled using the WLES $S$-$\Omega$ model. The SST $k$-$\omega$ model was used for the RANS component of the flow. Our rationale for selecting the SBES model is based on the following:
\begin{itemize}
    \item The SBES model avoids using very fine mesh at the wall (impractical for accurate LES) or other approximations but instead uses a highly developed and validated RANS approach.
    \item With the SBES method, RANS models are used to generate an eddy viscosity in the wall regions which are zones of low interest (internal nozzle flow in case of nasal spray). In regions of fine mesh, where accurate transient data are required, it uses an LES model (i.e., the near nozzle region where liquid sheet formation and disintegration occur). 
    \item The eddy viscosity from the RANS model and subgrid viscosity are blended with a fast transition which is very relevant in spray modelling.
    \item The SBES models shielding properties are much improved relative to Detached Eddy Simulation/Delayed Detached Eddy Simulation model formulations.
\end{itemize}

\subsubsection{VOF-DPM transition modelling}
The VOF to DPM multiphase CFD model in Ansys Fluent 2020R1 was employed to simulate atomization involving the liquid sheet break up into discrete droplets. This modelling approach allows the initial jet and primary breakup to be captured using the VOF model using refined mesh, whereas the disperse phase is modelled using Discrete Particle Modelling (DPM). This model reduces the computational expense of using only the VOF method which requires capturing of the liquid phase, its break-up into droplets (disperse phase) and their subsequent motion. The algorithm used in the VOF-DPM model accesses the eligibility for the transition of the liquid phase to the disperse phase based on two criteria: a) liquid lump size, and b) asphericity. 

This current study converted a lump size range of 5-80 $\mu$m into DPM droplets. Liquid droplets found in the VOF solution whose volume was greater than 80 $\mu$m were not converted to Lagrangian droplet parcels but rather allowed to break up on their own. 

The second eligibility criteria (asphericity) used for lump transition was based on the upper limit of asphericity calculated using the radius standard deviation and the radius surface orthogonality. In order to segregate lumps based on the radius standard deviation for DPM transition, the distance between the facet centre and the lump centre of gravity was calculated for every facet of the lump surface (gas-liquid phase interface). These distance values were weighted by the size of the individual lump boundary facet. The resulting standard deviation of these values was then calculated and normalized by the mean radius. This quantity is zero for a perfect sphere and increases to 1 as the shape deviates from spherical. In this study, lumps with a value less than 0.5 were transitioned to the discrete phase. Radius-surface orthogonality specifies the maximum asphericity calculated from the average radius-surface orthogonality. In order to calculate this quantity, a vector from the lump's centre of gravity to the centre of the lump boundary facet was computed for every facet of the lump surface. This vector was normalized and then used in a dot product with the normal vector to the mesh face, to yield a measure of relative orthogonality ranging from 0 to 1. For a perfect sphere, the value would be 1. A facet area-weighted average of these values was then subtracted from 1 to obtain the asphericity value. For almost spherical droplets, this asphericity value is much closer to 0 than the standard deviation-based asphericity. In the case of highly non-spherical lumps, the values generated by this method are similar to those obtained using the asphericity based on standard deviation. A threshold of 0.5 was used for the radius-surface orthogonality limit in the current study.

The phase transition process of converting liquid lumps to Lagrangian parcels does not cause volume displacement in the continuous phase. However, this approach does not conserve volume, which can result in momentum source errors. Therefore, to avoid spurious momentum sources, the VOF-DPF algorithm creates a gas volume with the same volume as that of the liquid lump to conserve volume. The mass source that is equivalent to the mass of the gas volume will affect the overall mass balance.

\subsubsection{Discrete Phase Modelling (DPM)}
The lumps that satisfy the transition criteria were converted to droplets and tracked using the Lagrangian discrete phase modelling approach to predict the particle trajectory. The trajectory of a discrete phase particle was predicted by integrating the force balance on the particle which includes drag, inertia, and gravity:

\begin{equation}
    m_p\frac{d \vec{u}_p}{dt}=m_p\frac{\vec{u}-\vec{u}_p}{\tau_r}+m_p\frac{\vec{g}\left(\rho_p-\rho \right)}{\rho_p}
\end{equation}

where $m_p$ is the particle mass, $\vec{u}$ is the fluid phase velocity, $\vec{u}_p$ is the particle velocity, $\rho$ is the fluid density, $\rho_p$ is the density of the particle,  $m_p(\vec{u}-\vec{u}_p)/{\tau_r}$ is the drag force, $\tau_r$ is
the droplet relaxation time calculated by
\begin{equation}
    \tau_r=\frac{\rho_p d^2_p}{18 \mu}\frac{24}{C_d \mbox{Re}}
\end{equation}
where, $\mu$ is the molecular viscosity of the fluid, $d_p$ is the particle diameter, and $\rm Re$ is the particle Reynolds number, which is defined as 
\begin{equation}
    {\rm Re} = \frac{\rho d_p \vert \vec{u_p}-\vec{u} \vert}{\mu}
\end{equation}

As a droplet is distorted from a spherical shape, the drag coefficient changes. The drag coefficient is calculated using a dynamic drag law theory \citep{Liu1993}. For spherical droplets, the drag coefficient is given by

\begin{equation}
  {C_d,_{sphere}} =
  \begin{cases}
    0.424 & \text{Re$\  >\  $1000} \\
    \frac{24}{\rm Re} (1+\frac{1}{6}{\rm Re}^\frac{2}{3}) & \text{Re$\ \leq\ $1000}
  \end{cases}
\end{equation}

As an initially spherical droplet moves through a gas, its shape is distorted if its Weber number is high enough to initiate the breakup.  In the extreme case, the droplet shape will approach that of a disk. During this process, the drag coefficient is given by

\begin{equation}
    C_d = {C_d,_{sphere}} (1+2.632y)
\end{equation}

where $y$ is the droplet distortion, as determined from the Taylor analogy break-up model (TAB) \citep{Taylor1963} used here. At $y=0$, the drag coefficient of a sphere is obtained, while at $y=1$, maximum distortion with the drag coefficient corresponding to a disk is obtained. This model is a widely used method for calculating droplet break-up for low Weber number applications based upon Taylor's analogy. The distorting and oscillating droplet subject to surface tension, drag and viscous forces are modelled by analogy with a spring-mass system representing the restoring force of the spring, an external force and a damping force, respectively. When the droplet oscillations grow to a critical amplitude the ``parent" droplet breaks up into a number of smaller ``child" droplets.

Two-way coupling between the discrete phase and the continuous phase was used to account for the interaction between the two phases. 

\subsection{Numerical Method and Simulations}
\subsubsection{Geometry}
\begin{figure} [h!]
	\centering
	\includegraphics[width=0.85\linewidth]{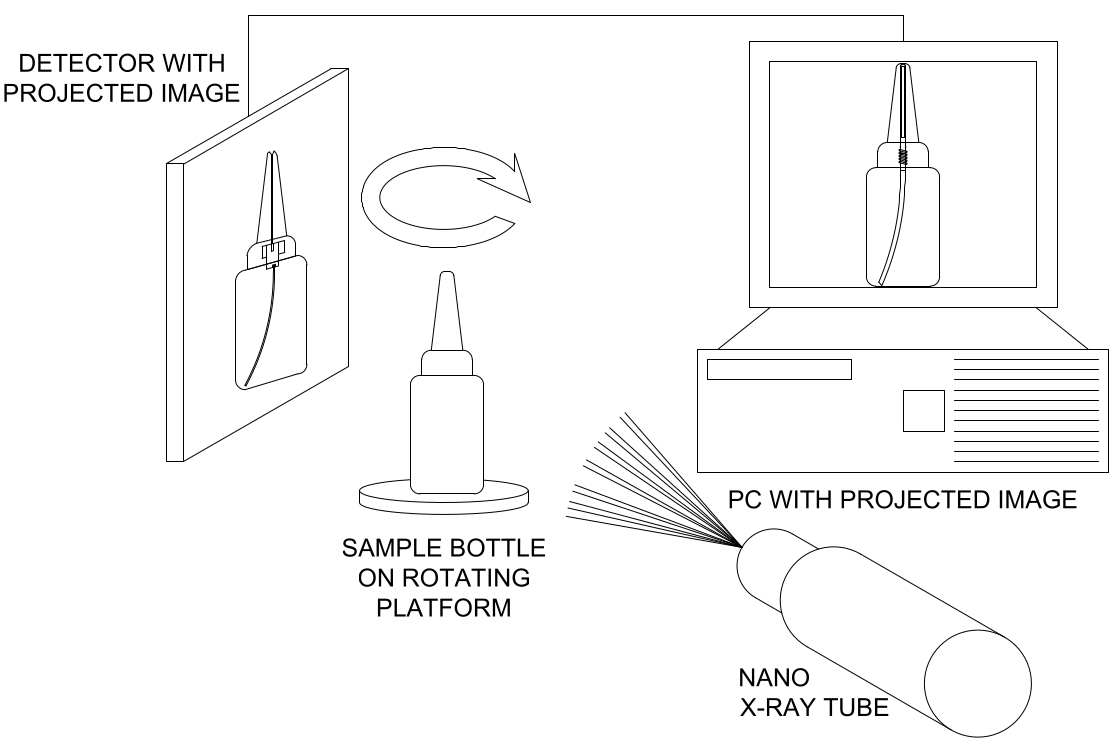}
	\caption{Schematic of the X-ray scanning method.}
	\label{fig:XrayfilmSchematic}
\end{figure}

\begin{figure}[h]
	\centering
	\includegraphics[width=0.8\linewidth]{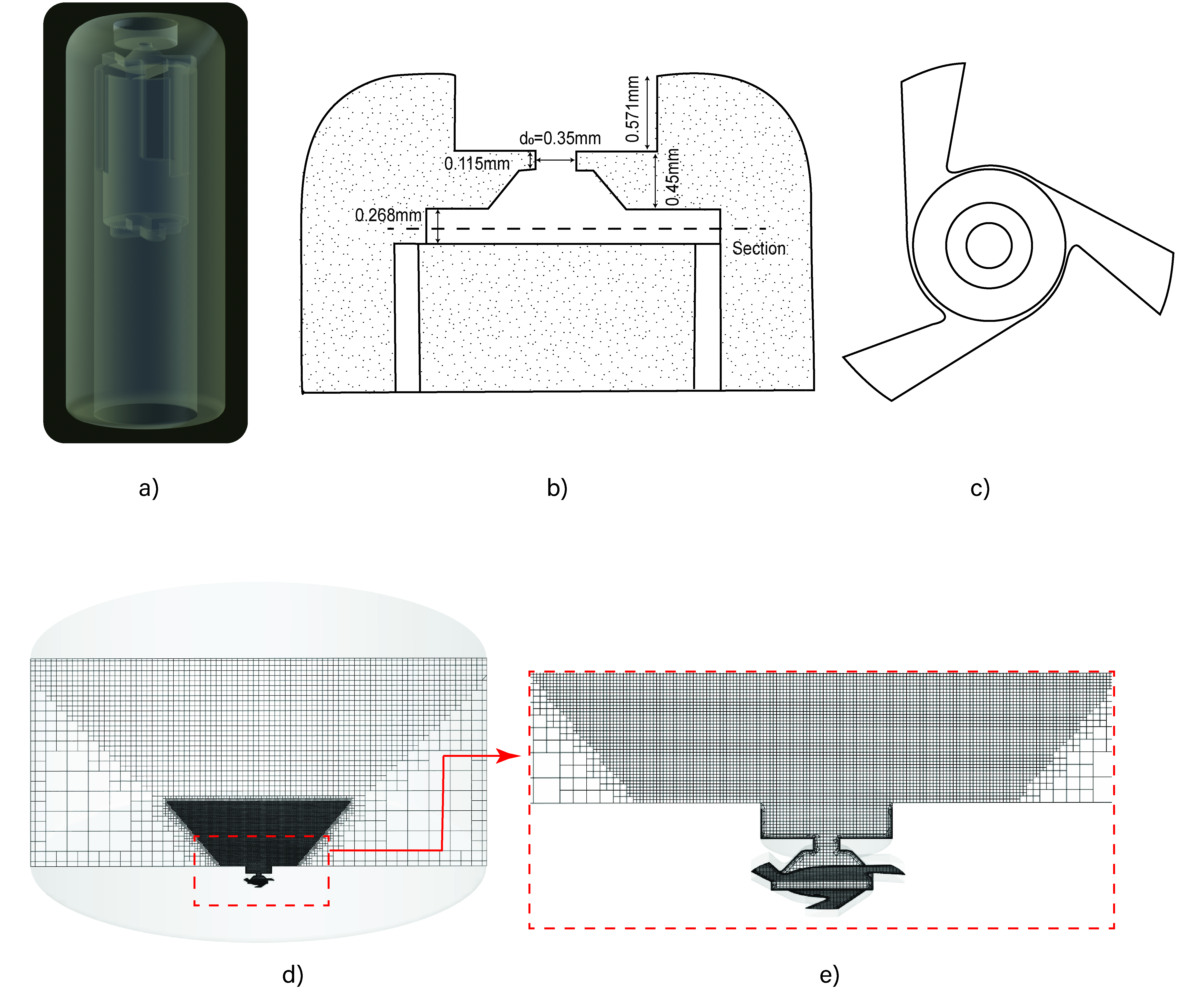}
	\caption{a) Rendered x-ray view of the nozzle internals, b) Vertical cross-section across the center of the nozzle, c) Horizontal cross-section across the swirl inlet port, d) Computational domain and e) Mesh resolution at the nozzle internals.}
	\label{fig:Geometry}
\end{figure}

The internal features of the Cophenylcaine nasal spray were extracted using a 3D Nano X-ray scanner. The schematic of the experimental set-up of the X-ray unit and its accessories are illustrated in the Fig.~\ref{fig:XrayfilmSchematic}). The extracted geometry was reverse engineered to create a 3D CAD model (See Fig.~\ref{fig:Geometry}a). The geometry consisted of three inlet ports that drive the liquid tangentially into the swirl port and it is ejected out of the nozzle with a diameter of $\ d_0=0.35\,$mm. 

A vertical cross-section through the nozzle orifice and a horizontal cross-section through swirl ports of the nasal spray illustrating its internal features are shown in Fig.~\ref{fig:Geometry}b and \ref{fig:Geometry}c, respectively. A cylinder was attached to the nozzle exit whose diameter and height were $d_c=85d_0$ and $h_c=42d_0$, respectively, to capture the primary and secondary break-up. A 3D computational domain is depicted in Figs~\ref{fig:Geometry}d and \ref{fig:Geometry}e. Three planes were created at $8\,$mm, $12\,$mm and $15\,$mm from the nozzle tip to collect droplet samples to evaluate the droplet size distribution.

\subsubsection{Mesh and Adaptive Mesh Refinement (AMR)}
Cut cell meshing was chosen to minimize the cell count and keep the aspect ratio low for LES with an initial mesh count of 3.6 million. The nozzle geometry and near-nozzle primary break-up regions were refined using a body of influence local refinement ($\Delta x=0.165\,$mm). The AMR technique was used to refine the cells in the liquid-gas interface to capture the primary break-up and turbulent structues in the flow. Three levels of mesh refinement were used, which could capture the primary break-up sufficiently with the lowest level mesh size of 5.5 microns. 

Cell registers were used to refine and coarsen the cells based on two criteria a) iso-surface value and b) volume fraction curvature. The cells whose volume fractions were greater than 0.55 and had a VOF interface with curvature more than \num{1e-12}~m$^{-1}$ were categorized for refinement, whereas the cells whose volume fractions were smaller than 0.45 and were away from the liquid-air interface with curvature less than \num{1e-14}~m$^{-1}$ were categorized for coarsening. The curvature method captures the largest possible range of cells on both sides of the phase boundary (liquid and gas). The final cell count during the stable phase of spray development was in the range of 7-7.5 million.

\subsubsection{Initial and Boundary Conditions}
To replicate the effect of nasal spray priming, the nasal spray device was assumed to be filled with liquid to the nozzle throat. This region of the domain was initialized with the volume fraction of liquid as 1 ($\alpha_l=1$). The physical properties of water were used to represent the nasal solution and ambient air was used as the gas phase. The liquid was initialized with a uniform velocity of $5\,$m/s at the inlet ports.

\subsubsection{Solver Settings}
 The PISO solver was used to couple velocity and pressure since it provides good solution stability. The Least Squares Cell-based method was used for the gradient discretization and PRESTO! was selected for the pressure spatial discretization in order to get better accuracy. Bounded central differencing was used for the momentum discretization and geo-reconstruct for the volume fraction. An interfacial anti-diffusion treatment was applied to limit the numerical diffusion that can arise from the volume fraction advection schemes. Second-order upwind was used for the turbulence discretization and first-order implicit for the transient formulation, as this is a requirement when using explict interface tracking.

The transient flow was modelled with a variable time step restricted by a Courant number of 1. The minimum and maximum time step was set to 10$^{-9}$~s and 10$^{-6}$~s, respectively, and the timestep was adjusted based on the smallest cell size and the highest velocity in the domain.
\
\section{Results}
\subsection{Pre-Atomisation Verification}
\begin{figure}[h!]
	\centering
	\includegraphics[width=0.9\linewidth]{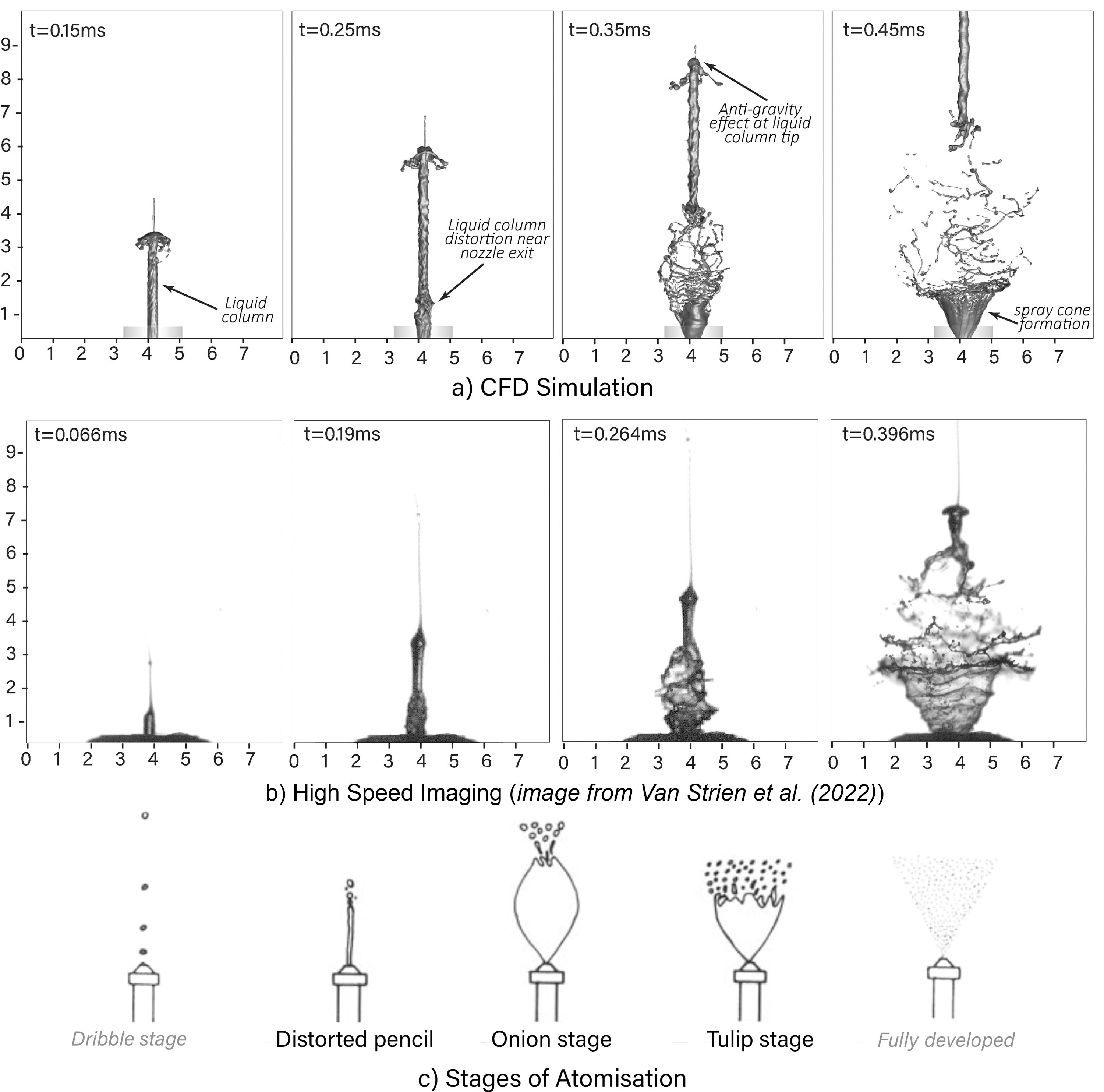}
	\caption{Pre-atomisation stage results from  (a) CFD simulation and (b) High Speed Imaging, demonstrating a swirling liquid column transition into a conical liquid sheet reported in \cite{Van2022}. The time steps for the CFD Simulation and High Speed Imaging are slightly offset, however, the duration of the early period remains the same for both at $\Delta t = 0.3$~ms. (c) The spray development stages of atomisation are described in  \cite{lefebvre2017atomization} schematically as distorted pencil, onion stage, tulip stage, and fully developed.}
	\label{fig:Earlyphase}
\end{figure}

Figure \ref{fig:Earlyphase} shows a qualitative comparison of the simulation results against the experiment imaging (reported in \cite{Van2022}) at an early phase of spray development. A time lag (difference) between the CFD simulation and high-speed imaging was observed (Figure \ref{fig:Earlyphase}). 

The inconsistency in time was due to the difference between the real actuation force used in the experimental visualisation and the constant inlet flow velocity used in the CFD simulation. While the constant inlet velocity condition in the CFD model does not represent the step profile of the real spray actuation force exactly, the majority of the spray event is consistent, and the effects due to the difference are evident in the time difference found in the panels of Figure \ref{fig:Earlyphase}.

An initial cylindrical liquid column jet is ejected from the nozzle. The column jet then expands and becomes distorted close to the nozzle exit as instabilities begin to form due to the centrifugal forces. As the mass flow increases, the liquid column distortion  intensifies and eventually developes into a conical-shaped liquid sheet structure, where Kelvin-Helmholtz waves appear on the liquid sheet to produce liquid sheet instabilities. There is a competing interplay of inertial and centrifugal force provided by the upstream pressure to expand and eventually break up the sheet, while the viscous and surface tension forces of the liquid aim to hold the sheet together. The five stages of atomisation described by \cite{lefebvre2017atomization} are shown in Figure \ref{fig:Earlyphase}c. Our results exhibit the distorted pencil, onion, and tulip stages in the early phase of atomisation. 

 \begin{figure}[h]
	\centering
	\includegraphics[width=0.65\linewidth]{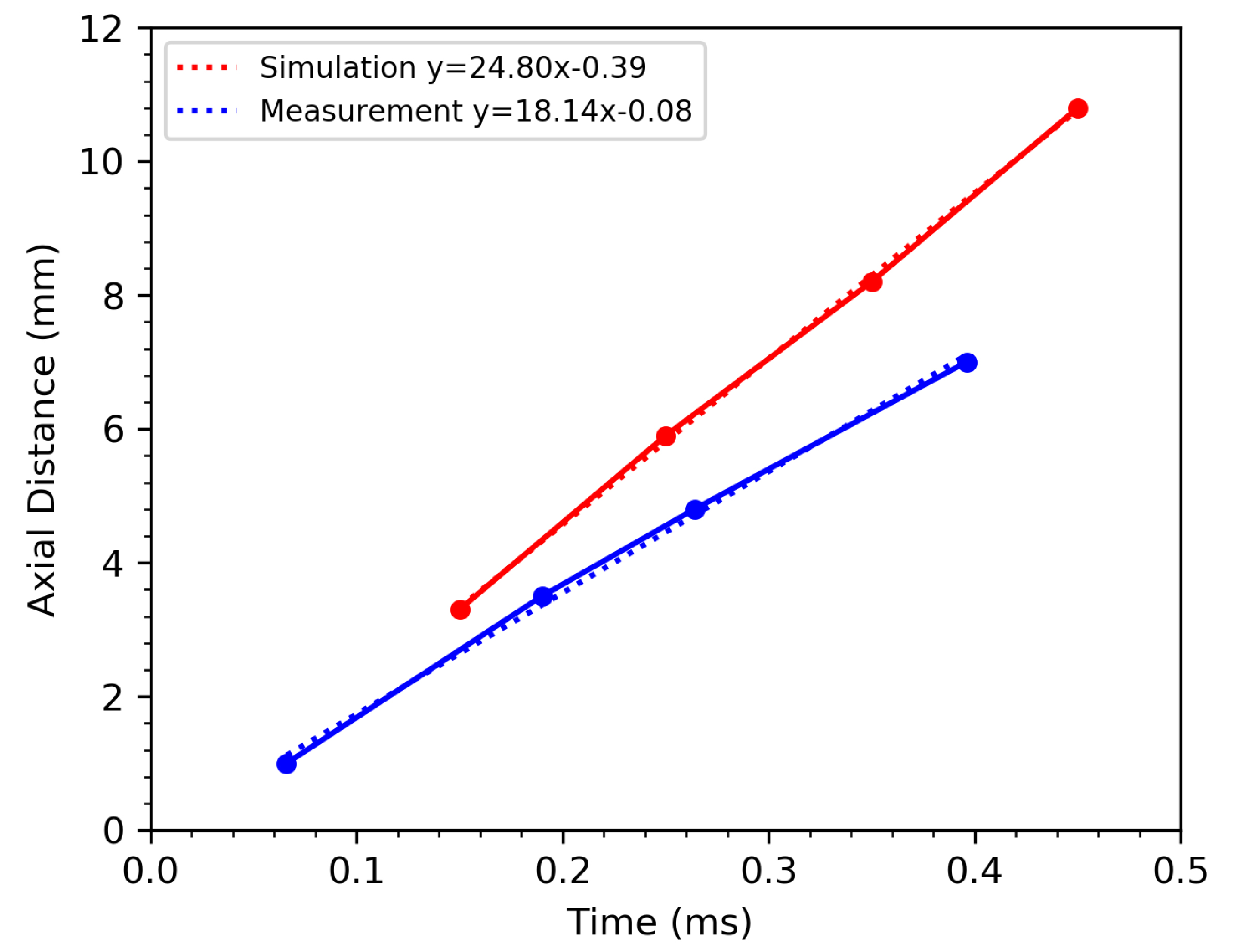}
	\caption{Comparision of liquid penetration at an early stage of spray development of a measured \cite{Van2022} and simulated liquid column. They are fitted to a linear regression line whose slope determines the penetration velocity magnitude.}
	\label{fig:Penetration}
\end{figure}

The spray penetration during the early phase of spray development is given in Figure~\ref{fig:Penetration}. The slope of the line determines the liquid jet velocity where the CFD simulation results showed an average velocity of 24.80 m/s, which was higher than the measurement data which had 18.14 m/s. In general, the CFD simulation agrees well with the experimental result qualitatively, if we account for the time lag difference between the CFD simulation and measurement results.

\subsection{Liquid Sheet Breakup}
\begin{figure}
	\centering
	\includegraphics[width=1\linewidth]{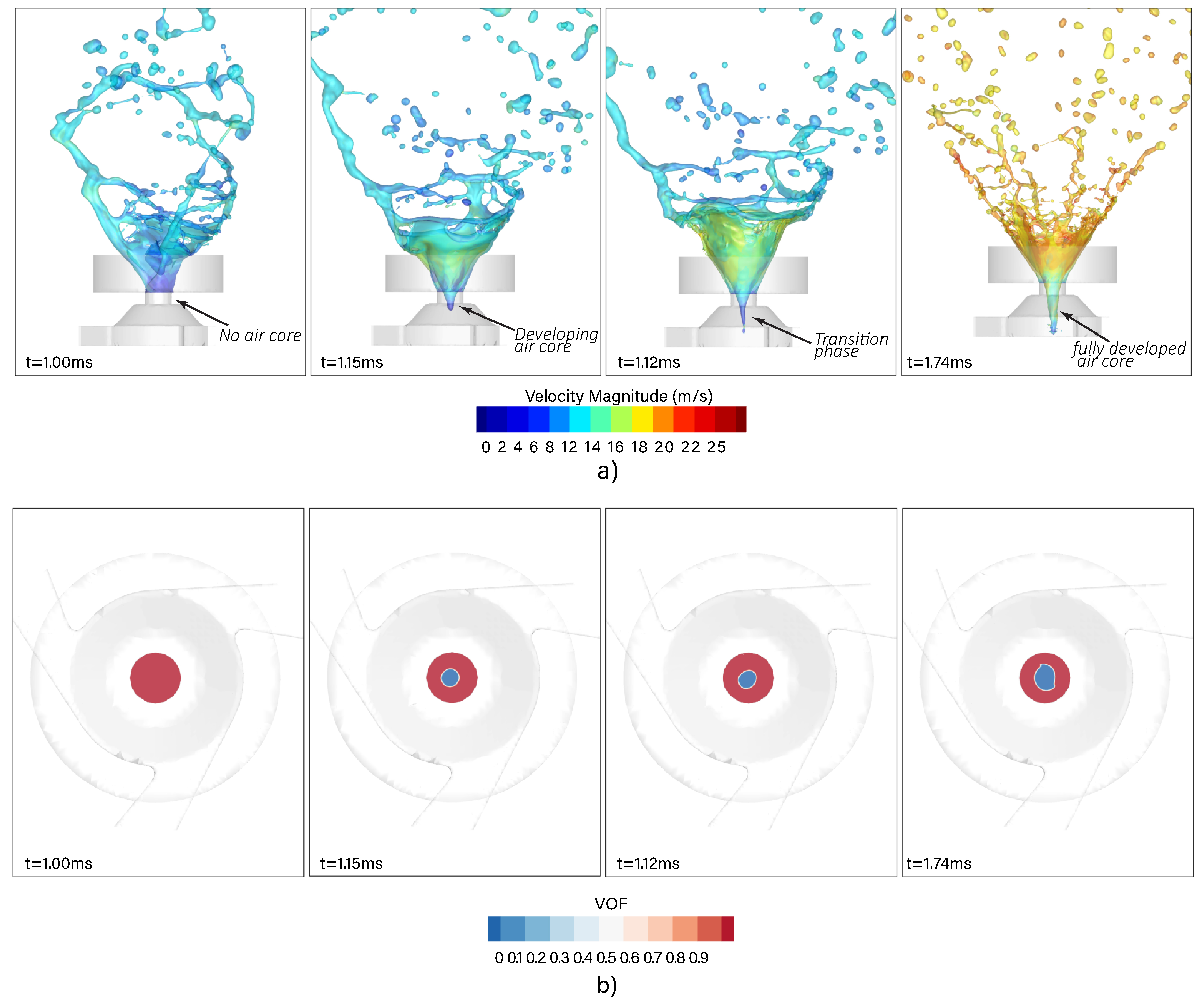}
	\caption{a) Air-core formation during early stage of spray development depicted with iso-surface of VOF at 0.5 colored with velocity magnitude and b) Section plot of VOF at nozzle throat depicting the air-core diameter during the early stage of spray development. The diameter of the air core fluctuates between 0.13-0.20\,mm during the fully developed phase.}
	\label{fig:Aircore}
\end{figure}

During the main atomisation phase, a characteristic air-core region is formed in the nozzle, as shown in Figure \ref{fig:Aircore}. The initial conditions of the simulation defined the swirl chamber and the nozzle throat domain with liquid to represent a primed nasal spray device. As the simulation proceeded, the liquid moved through the swirl chamber and exited the orifice with a tangential velocity component resulting from the tangential ports. A negative pressure was created in the orifice centre, and an air-core was formed ($t=1.15$ ms in Figure \ref{fig:Aircore}). 

The fully developed stage of the spray can be identified by its maximum air-core diameter. The air-core diameter is shown at different stages of the spray formation represented by the liquid volume fraction (blue colour) in Figure \ref{fig:Aircore}b. The air-core region exists during the entire stable phase of the spray development and fluctuates in shape and size during the atomisation as the liquid sheet is formed from the pressure swirl atomizer.

\begin{figure}
	\centering
	\includegraphics[width=1\linewidth]{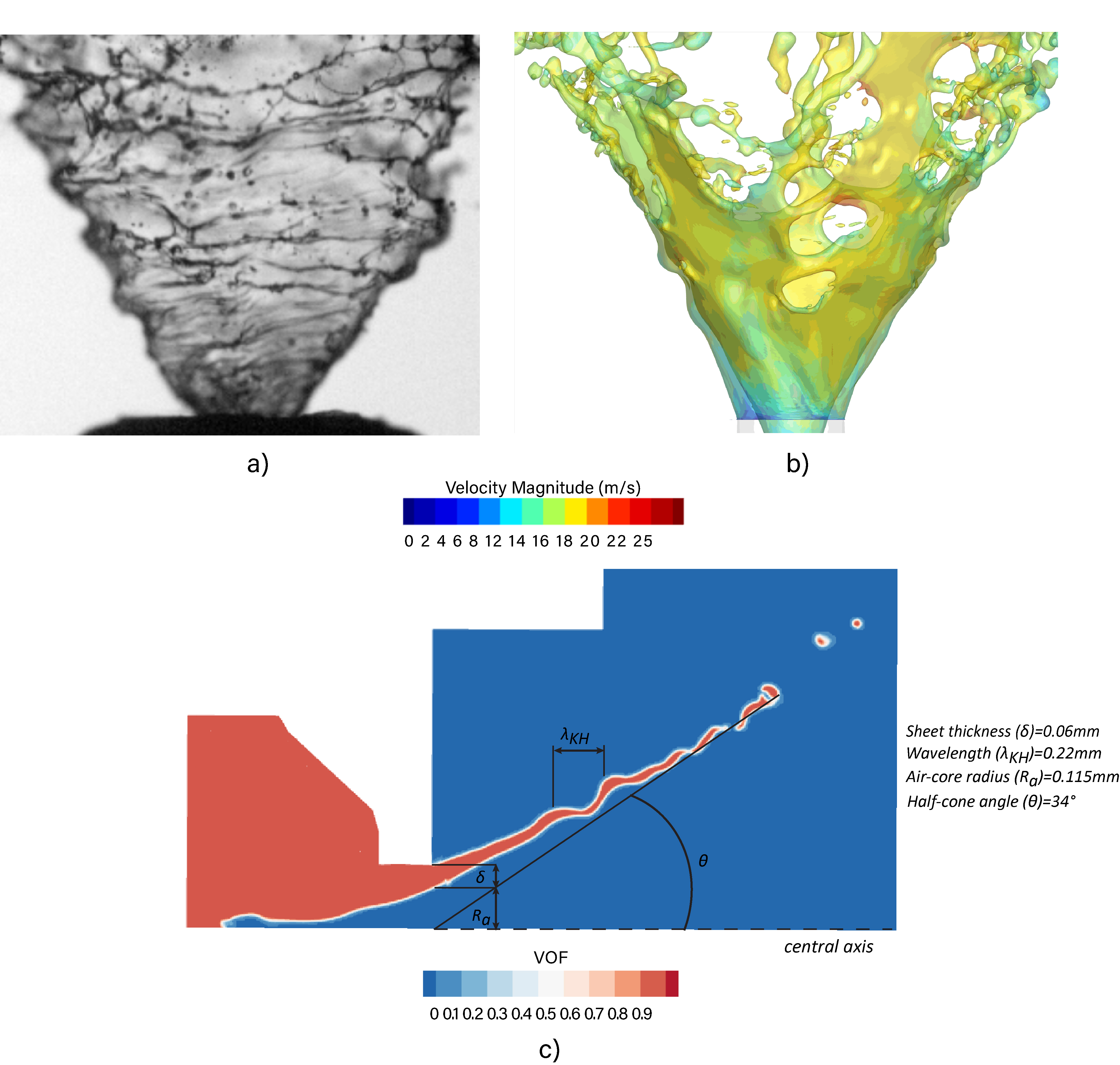}
	\caption{Near nozzle liquid sheet depicting the surface wave formation from a) an experiment \cite{Van2022} and b) a simulation and c) Primary Break-up depicting liquid sheet thickness, air-core radius, KH wave development and spray half-cone angle.}
	\label{fig:KH_Instability}
\end{figure}

In the near nozzle region, Kelvin-Helmholtz (KH) instabilities produce waves on the liquid sheet shown in Figure \ref{fig:KH_Instability}a, b and those with the wavelength for maximum growth cause periodic thickening of the liquid sheet in a direction normal to the flow \cite{lefebvre2017atomization}. The length scale of the KH instabilities measured in the experiment was similar to that in the simulation. The distance between the two wave amplitudes was defined by the wavelength. 

Columnar ligaments break off from the thinning conical sheet, where the liquid volume contained in the ligament can be estimated as the volume of the ligament cut out of the sheet with a thickness equal to that of the sheet at the breakup distance and a width equal to one wavelength, shown in Figure \ref{fig:KH_Instability}c, through the volume fraction of the liquid at the central axis of the jet. These cylindrical ligaments disintegrate into droplets according to the Rayleigh mechanism \citep{york1953mechanism}. The parameters such as sheet thickness, wavelength of the disturbance, air-core radius and half-cone angle are depicted in the figure. The spray cone angle and the breakup length are the most important parameters for primary break-up which characterizes the spatial distribution range of the liquid film, while the latter characterizes the break-up position of the liquid film from the nozzle exit.

\begin{figure}
	\centering
	\includegraphics[width=1\linewidth]{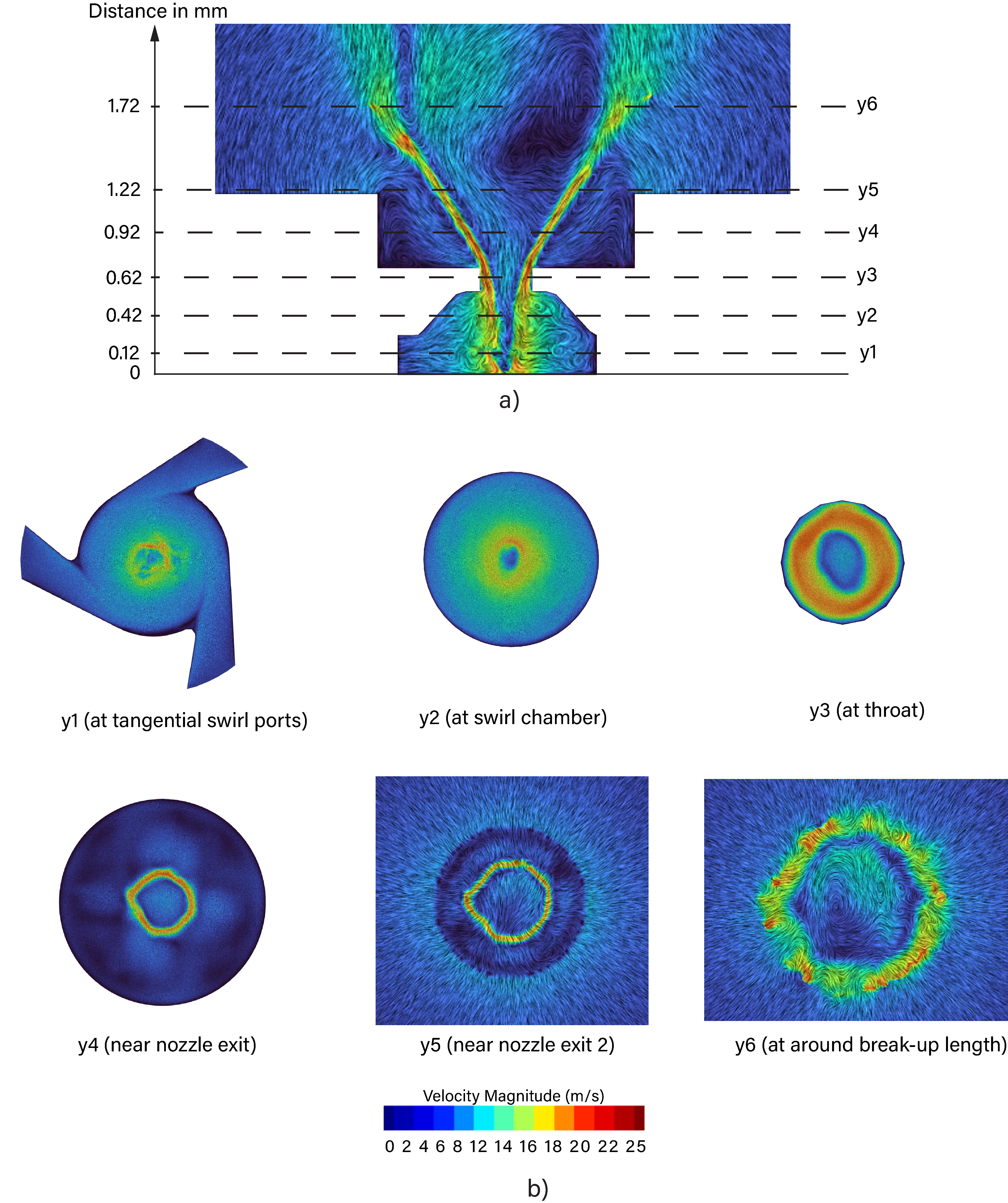}
	\caption{a) Vertical slices through the centre of the nozzle and b) Horizontal slices at y1, y2, y3, y4, y5 and y6 depicted in figure a).}
	\label{fig:Velocity_Lic}
\end{figure}

Figure \ref{fig:Velocity_Lic}a illustrates the line integral convolution of the velocity flow field which visualizes the coherent flow structures along the vertical and horizontal sections through the central nozzle axis.
In Fig.~\ref{fig:Velocity_Lic}b, sections y1-y3 shows the internal nozzle flow-field, whereas sections y4-y6 shows the near nozzle flow-field where the interaction of the liquid with the static gas is observed.  The swirling liquid jet disintegration is classified in $We$-$MR$ space \citep{Rajamanickam2017}). Here, $We$ represents the Weber number and $MR$ represents the momentum ratio (liquid-to-gas momentum ratio). In a typical nasal application, inertial force are dominant over the surface tension force with $We<2$ and $MR=0$ (with the assumption that gas is quiescent). This implies, nasal spray lies in an axisymmetric regime in We-MR space generating axisymmetric spray, as illustrated in Fig.~\ref{fig:Velocity_Lic}a.
\begin{figure}
	\centering
	\includegraphics[width=1\linewidth]{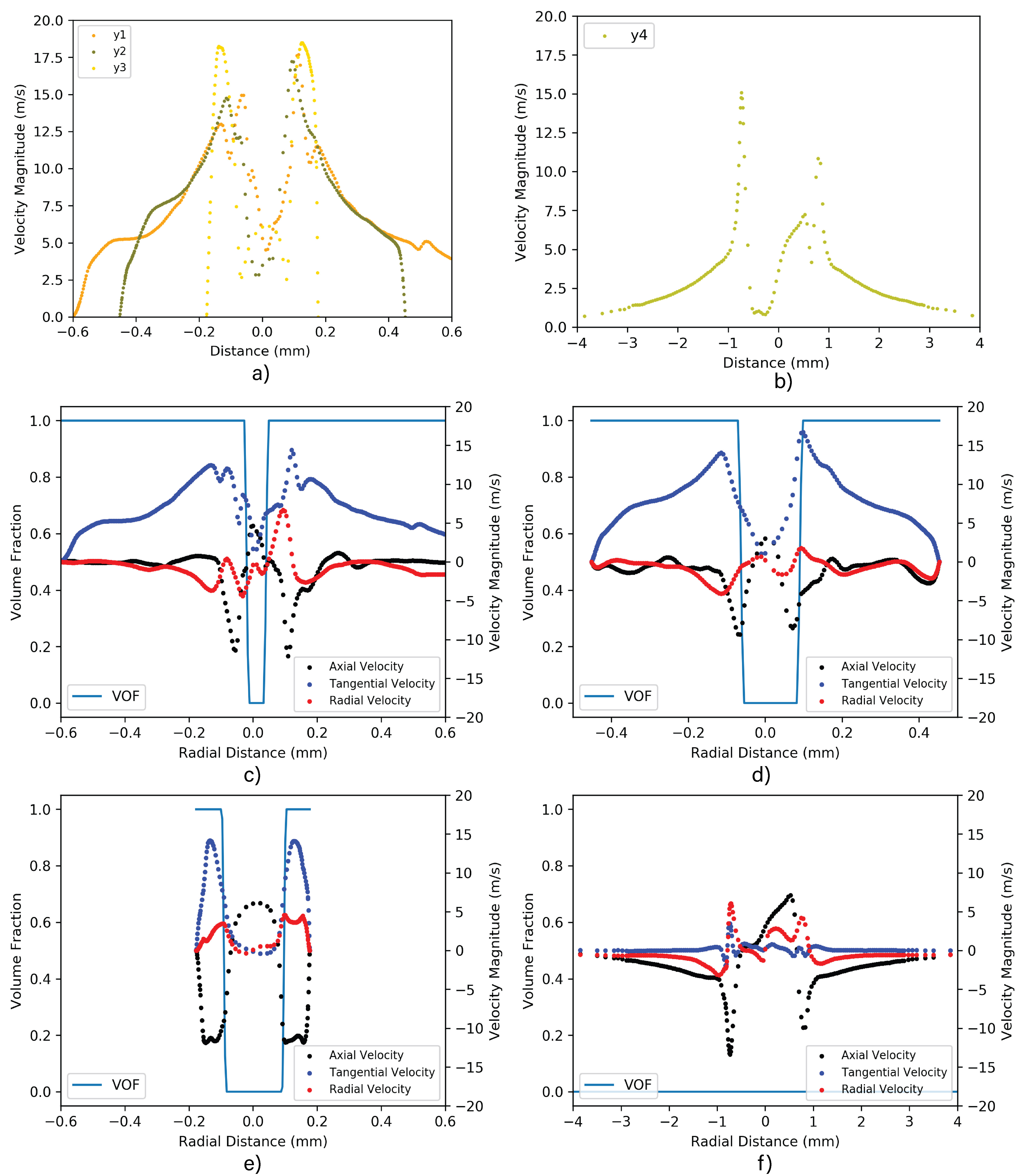}
	\caption{Velocity magnitude a) at the nozzle internals at section y1, y2 and y3 and b) at the break-up length. Velocity components and volume fractions of water c) at tangential port (y1), d) at swirl chamber (y2), e) at the nozzle throat (y3) and f) at the breakup length (y4).}
	\label{fig:Vel_Prof_all}
\end{figure}
Figure~\ref{fig:Vel_Prof_all}a and b depict the velocity magnitude profiles along the mid-lines taken at locations y1 to y4, illustrated in Fig.~\ref{fig:Velocity_Lic}. The velocity magnitude was decomposed into its axial, tangential, and radial components are shown in Figure \ref{fig:Vel_Prof_all}c-f.   The velocity profile in the nozzle showed a bimodal distribution. Along the tangential port and swirl chamber cross-section, the tangential velocity of the liquid was much larger than the axial and radial velocities indicating that the spiral swirl motion of the liquid is dominant. At the nozzle throat, the tangential and the axial velocity were of similar magnitude with their peak values at the air-liquid interface. The peak radial velocity component was one-third of the other two components. The extent of the volume fraction in the figure signifies air-core vortices at different sections of the spray atomizer. At the break-up length, the axial velocity magnitude was maintained remaining dominant, however the tangential velocity diminished significantly. 

\begin{figure}[h]
	\centering
	\includegraphics[width=1\linewidth]{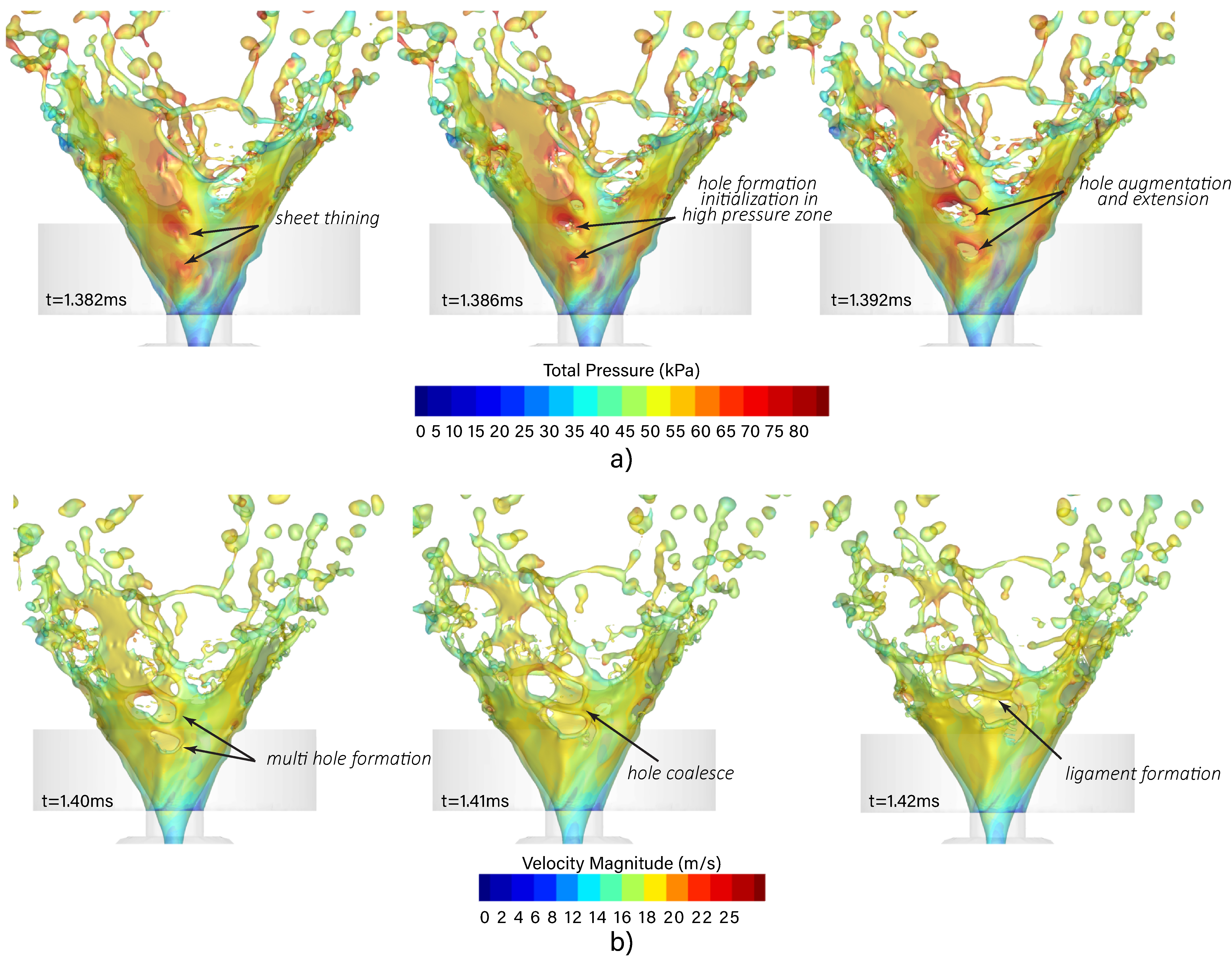}
	\caption{a) Liquid sheet hole formation at high-pressure zone and b) Liquid sheet undergoing hole extension and ligament formation.}
	\label{fig:Ligament}
\end{figure}

Figure \ref{fig:Ligament}a shows the bulk liquid sheet, coloured by the total pressure, at the fully developed phase of the spray formation. The liquid sheet thinning process was identified at the high-pressure regions. The continued thinning of the liquid sheet led to sheet perforation (holes in the sheet) under the action of inertial and centrifugal forces. The competing viscous and surface tension forces would act to resist the thinning to prevent the sheet break-up. 
Figure~\ref{fig:Ligament}b shows that as multiple perforations appear, thin ligaments are produced due to the coalescence of the perforations (e.g., holes).

\begin{figure}
	\centering
	\includegraphics[width=1\linewidth]{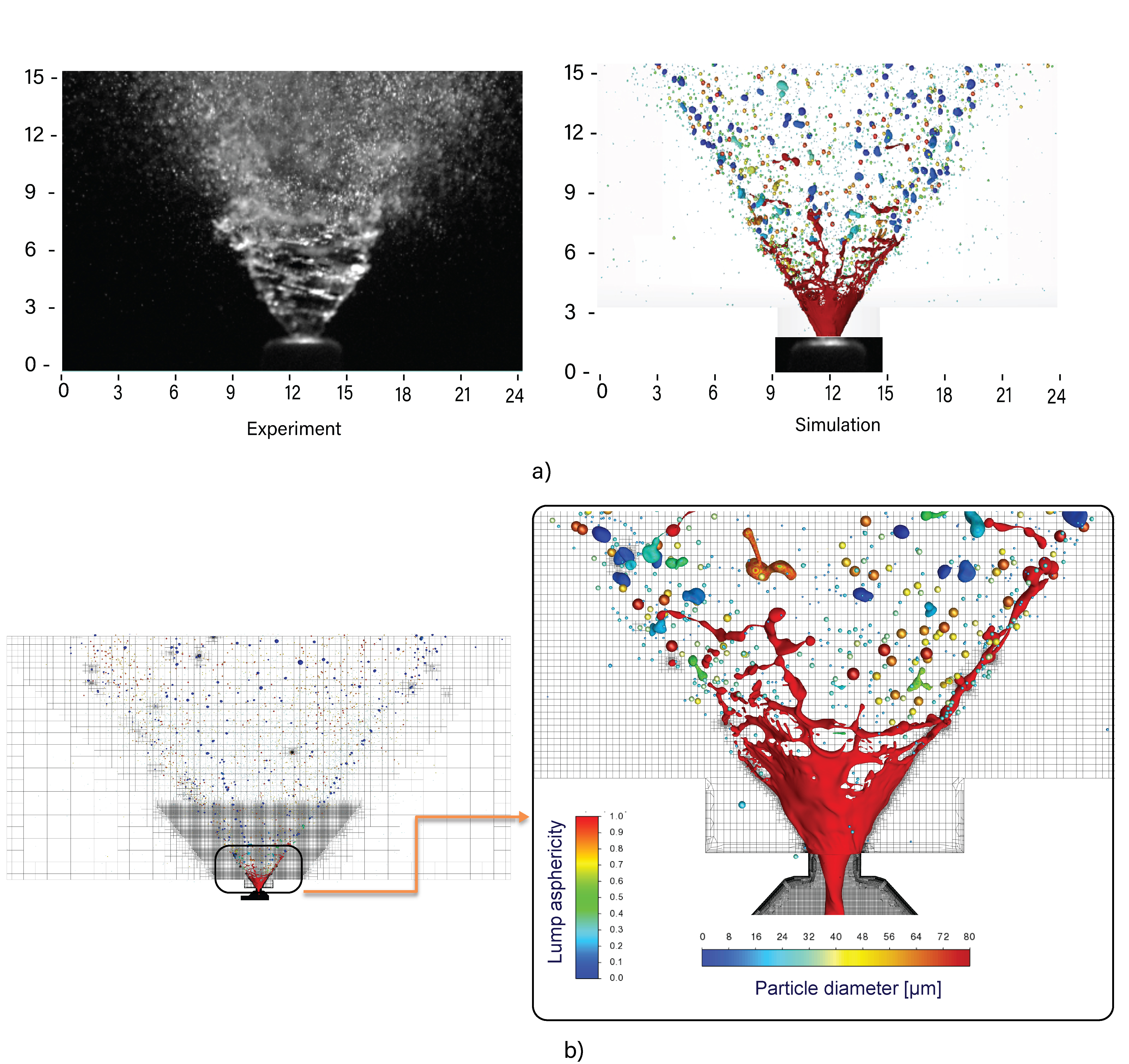}
	\caption{a) Qualitative comparison of a fully-developed spray at the stable stage from an experiement \cite{Van2022} and simulation b) VOF to DPM conversion of liquid lumps based on lump asphericity criteria. Lumps with asphericity $< 0.5$ are converted into discrete particles.}
	\label{fig:Full_spray}
\end{figure}

Figure \ref{fig:Full_spray}a depicts a qualitative comparison of the fully-developed spray at the stable stage of spray formation from the CFD simulation and experiment. Qualitatively, the spray plume volume and spray angle are well captured in the numerical simulation. Figure~\ref{fig:Full_spray}b shows the liquid lumps that are converted into discrete particles based on the lump asphericity criteria. The liquid lumps whose asphericity was less than 0.5 were converted into discrete droplets. The simulation resolved 98.99\% of the total spray mass, whereas the rest of the spray mass was diffused. 

\begin{figure}
	\centering
	\includegraphics[width=1\linewidth]{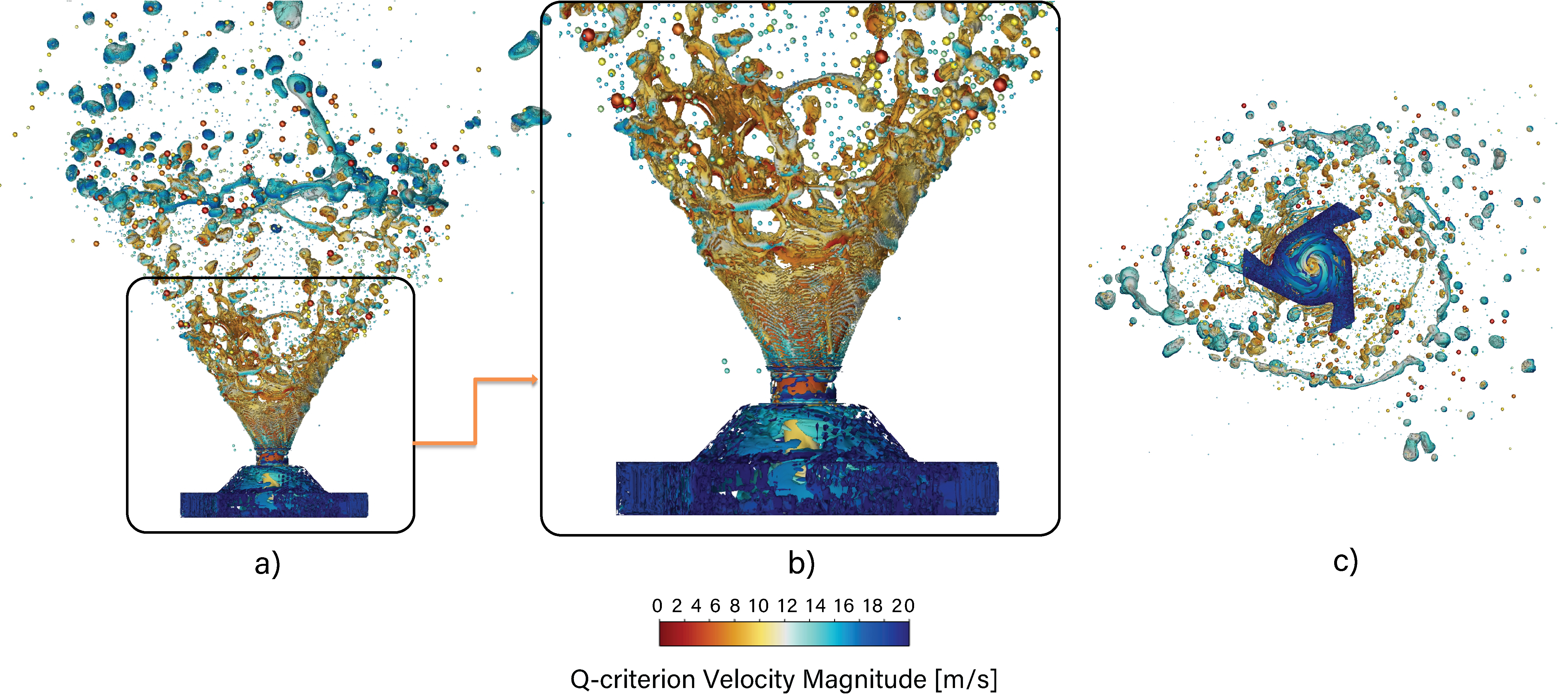}
	\caption{a) Q-criterion in liquid phase as iso-surface b) Magnified view of the near nozzle large scale turbulence and c) Bottom view }
	\label{fig:Q-criterion}
\end{figure}

\begin{figure}
	\centering
	\includegraphics[width=1\linewidth]{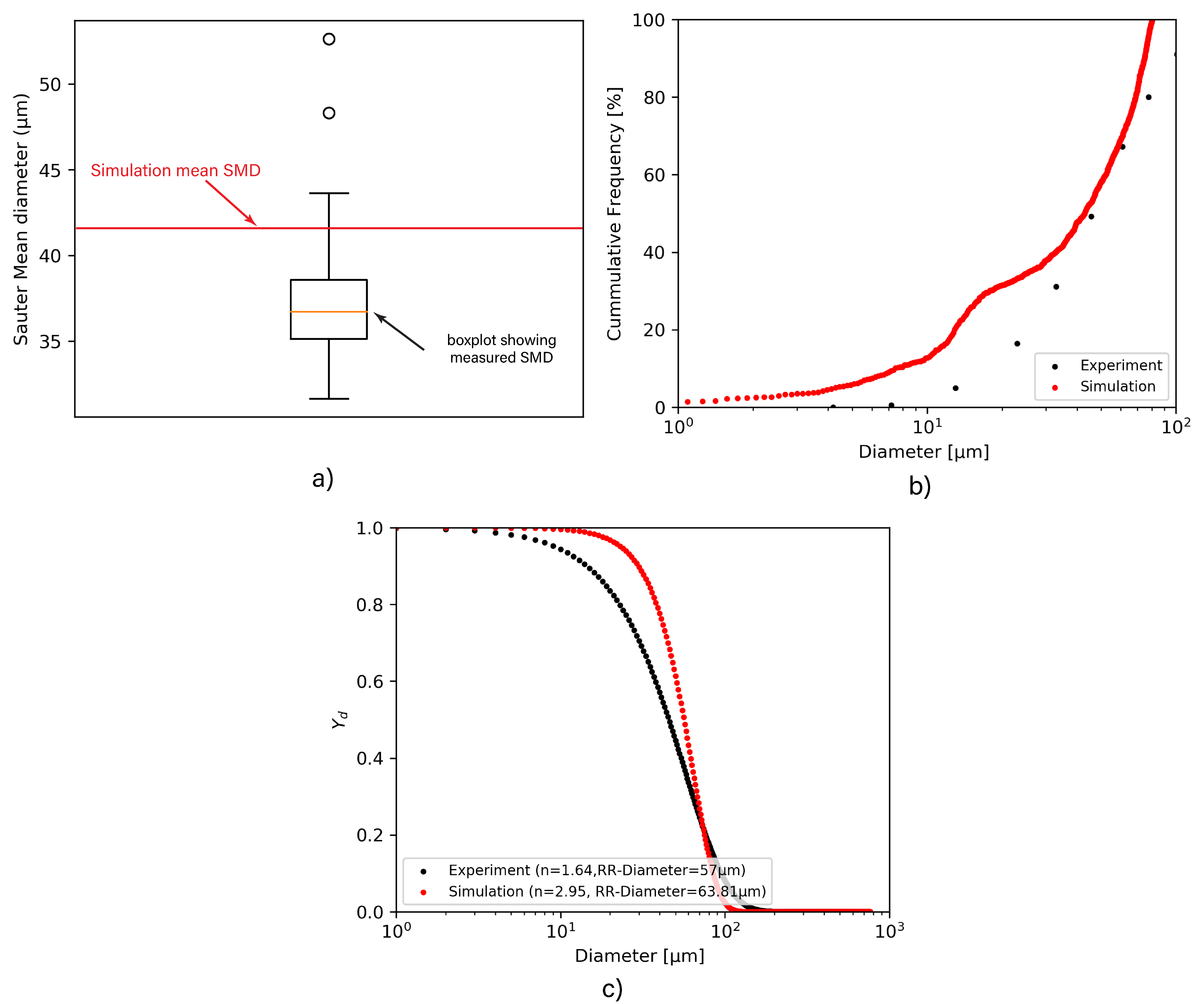}
	\caption{a) Boxplot of the measured SMD against simulated mean SMD, b) Droplet size distribution for measurement data against simulation results and c) Rosin-Rammler distributions for measurement and simulation. The quantitative experimental data are extracted from \cite{Van2022}.}
	\label{fig:DSD}
\end{figure}

Turbulence generated from fast moving liquid sheet coupled to the air phase is demonstrated in Figure \ref{fig:Q-criterion} via an iso-surface of the Q-criterion with a value of 1000~s$^{-2}$ in liquid phase coloured by velocity magnitude. The large-scale turbulence structures are resolved which can be observed in the near nozzle region and inside the atomizer.
The resulting atomised droplet diameter was calculated using the Sauter mean diameter. Figure \ref{fig:DSD}a shows a boxplot of the SMD measured in the experimental work against the simulated mean SMD. The mean SMD is over-predicted in the simulation, however, it was within the measured upper quartile. Figure \ref{fig:DSD}b shows the cumulative droplet size distribution (geometric diameter) of the measurement and the simulation and Figure \ref{fig:DSD}c shows the Rosin-Rammler distribution of the measurement against the simulation. The larger diameter particles ($>50\,\mu$m) were well-predicted while the smaller particles were found to be over-predicted. There are two possible reasons that account for the discrepancies between the experiment and numerical analysis. a) The experiment uses the spray diffraction method to obtain DSD that has a cut-off limit in detecting smaller droplets (i.e. $<10~\mu$m) with high accuracy. It is possible that these smaller droplets travelling at high velocity were not captured sufficiently. b) The TAB break-up model with the default break-up parcel was employed for the secondary breakup. This implies that when the TAB break-up criteria are satisfied, a parent particle generates only one child parcel and one parent parcel, resulting in a limited number of secondary break-up parcels. Consequently, smaller droplets tend to be over represented, resulting in the overpredicted smaller particles.
\section{Discussion}
Spray atomisation can be categorized into three phases on the basis of the near-nozzle spray plume shape, i) early pre-stable or expanding phase, ii) fully developed stable phase and iii) collapsing phase. 

During the pre-stable stage, before the spray is formed, a liquid column with relatively low swirling intensity is produced as it leaves the nozzle exit. At this stage, the aerodynamic force is relatively small with low axial velocity, with an absence of swirling motion, allowing the surface tension to dominate, limiting the deformation of the liquid jet. As the pressure increases from the liquid mass entering the atomizer, the higher pressure deforms the liquid (Figure \ref{fig:Earlyphase}).

A fully-developed stable stage of the spray is identified by the maximum spray cone angle and fully developed air-core formation in the atomizer. The air core is formed when the centrifugal force of the swirling flow overcomes the viscous force and a low-pressure area near the injector exit is created by the swirling liquid motion in the swirl chamber \citep{Amini2016}. Several parameters influence the diameter of the air core, and it increases in size with the actuation force \citep{Liu2017}. In nasal spray applications, the deviation in actuation pressure due to a patient's manual actuation of the device is minimal and varies between 0.2-0.6 MPa \citep{Van2022}, thus, its influence on the air-core diameter would be negligible. However, the internal nozzle geometry  influences the air-core diameter significantly; therefore, it impacts primary breakup and eventually influences droplet size distribution. The air core diameter decreases with an increase in the swirl chamber length, and as it reaches the critical length, the air core disappears \citep{Liu2017}. However, in nasal sprays, the swirl chambers are essentially short enough to retain the air core during the entire spray duration ($\sim$0.15\,s) for better controllability and repeatability of the doses.

Three distinct regions were identified during the fully-developed spray formation phase. Region 1 involves surface waves on the continuous liquid sheet jet; Region 2 has horizontal ligaments and; Region 3 is the zone of droplet formation. The liquid sheet breakup mechanism is a complex phenomenon because the flow pattern is much more complicated due to the interaction of several forces. Fraser et al. \citep{Fraser1953}  proposed three basic modes of liquid sheet disintegration a) Rim, b) Wave, and c) Perforated sheet. 

The rim mode is a prominent mode of disintegration for liquids with both high viscosity and surface tension \citep{Chinn1996}. This phenomenon was initially observed in the early stage of spray development when the surface tension force dominates (Figure \ref{fig:Earlyphase} at $t=0.45$~ms (Simulation) and $t=0.396$\,ms (Experiment)). When the pressure increases from the liquid filling the atomizer, the later modes of liquid sheet disintegration (wave, and perforated sheet) become prominent. In the rim mode of liquid sheet disintegration, surface tension forces cause the free edge of the liquid sheet to contract to form a thick rim. Due to the strong shear effect of the still gas and the liquid interface, the sheet rim becomes thinner to form thread-like structures, called  ligaments. These disintegrtae further to form droplets that continue to move in their original direction but remain attached to the sheet by a ligament. Both the droplets and the threads continue to disintegrate further. 

The second mode of liquid disintegration is due to the growth of the unstable wave at the interface between the gas and the liquid (Figure \ref{fig:KH_Instability}). These unstable waves, also called KH instabilities, produce fluctuations in velocity and pressure, which cause the surface wave (liquid sheet) to deform into ligaments. Typically, a nasal spray operates in the low $We$ number regime ($We$~=~1.06-2.71). $We$ provides the relative importance of the fluid inertia compared with the surface tension force which is given by $We=\rho \times U^2 \times L/\sigma$,  where $\rho$ is the gas density, $U$ is the average spray velocity of 18-30\,m/s, $L$ is the characteristic length taken as the orifice diameter of 0.35\,mm and $\sigma$ is the liquid surface tension. For nasal spray applications, $We < 2$ suggesting long-wave instabilities in the form of sinuous waves dominate.

In the perforated sheet breakup mode, holes appear in the high pressure zone of the liquid sheet (Figure \ref{fig:Ligament}a). The process starts with the sheet thinning under the action of the surface tension force. Later, these holes extend and merge, forming a series of net-like irregular ligaments.

Under the actions of the swirling motion and aerodynamic force, these liquid sheets and ligaments break into tiny liquid lumps. At this stage, a novel VOF-DPM transition method comes into play. The liquid lumps that meet the VOF to DPM conversion eligibility criteria based on asphericity and diameter, are converted into the discrete phase (Figure \ref{fig:Full_spray}). 

The current study has the following limitations.
\begin{itemize}
\item A constant velocity profile was used as an inlet boundary condition instead of the velocity profile represented by an actual hand actuation.
\item Detailed nozzle internal geometry was missing because the X-ray scan could not resolve geometric features such as fillets, rounds on the nozzle edges etc. This might lead to some changes in the flow dynamics of the spray.
\item The domain was limited to an axial length of 15\,mm to limit computational expense.
\item Mesh sensitivity study was beyond the scope of this research. However, to ensure high confidence in numerical accuracy, a very small time step in the range of $10^{-9}$ s to $10^{-6}$ s was employed (average time step was 10-8 s during the stable phase of spray development), along with three levels of adaptive mesh refinement, which limited the global Courant number to 1. The minimum grid size used was $5.5~\mu$m, which was approximately 11 times smaller than the liquid sheet thickness length scale.
\end{itemize}

\section{Conclusion}
Understanding the physical mechanisms of a breakup process of a swirling liquid out of a small-scaled pressure swirl atomizer for a nasal spray application is critical for controlled nasal drug delivery. Currently, numerical investigations available for nasal spray administration can only partially address the intricate physics associated with the primary atomization of a nasal spray. Hence, there is a need for computational investigations in the literature that can help understand the impact of the primary breakup of the liquid sheet on the secondary breakup of a nasal spray and the resultant droplet characteristics. With this motivation, the present work aims to investigate the internal nozzle flow and the formation of the liquid sheet, ligament and droplets in the swirling sheet breakup process using CFD. In this work, we reported numerical simulation using a novel VOF-DPM method using the adaptive mesh refinement (AMR) technique to resolve the liquid structure and co-related against the experiment. There are five stages of atomisation: dribble, distorted pencil, onion, tulip, and fully developed. Specifically, our results showcased the distorted pencil, onion, and tulip stages during the early phase of atomization. Furthermore, we observed that the nasal spray falls under the axisymmetric regime in the We-MR space, resulting in a more controllable and repeatable axisymmetric spray. Additionally, we identified three distinct modes of liquid sheet disintegration during nasal spray development: rim, wave, and perforated sheet. This study provided qualitative and quantitative data to better understand a nasal spray application's primary and secondary atomization. Qualitatively, the experimental results offered adequate evidence to the simulation result provided in the study

\section{Acknowledgements}
The authors gratefully acknowledge the financial support provided by the Garnett Passe and Rodney Williams Foundation Conjoint Grant 2019-22 and NCI Australia (National Computational Infrastructure) for providing with the high-end computing services.

\small
\bibliography{pressure}

\normalsize

\end{document}

%% file: preamble.tex
\usepackage[hidelinks]{hyperref}
\journal{Physics of Fluid}
\biboptions{authoryear}
\usepackage{amsmath}
\usepackage{caption}
\usepackage{subcaption}
\usepackage{graphicx}
\usepackage{upgreek}
\usepackage{hyperref}
\hypersetup{colorlinks  = true}
\usepackage{siunitx}
\usepackage{geometry}
\usepackage{array}

%% file: frontmatter.tex
\begin{frontmatter}
\title{Primary spray break-up from a nasal spray atomizer using  Volume of Fluid to Discrete Phase Model}


\author[label1]{Kendra Shrestha}
\author[label2]{James Van Strien}
\author[label3]{David F Fletcher}
\author[label1]{Kiao Inthavong \corref{cor1}}

\cortext[cor1]{Corresponding Author: kiao.inthavong@rmit.edu.au}
\address[label1]{Mechanical \& Automotive Engineering, School of Engineering, RMIT University, Bundoora, Victoria 3083, Australia}
\address[label2]{Faculty of Sci Eng \& Built Env, School of Engineering, Deakin University, Geelong Waurn Ponds, Australia}
\address[label3]{School of Chemical and Biomolecular Engineering, The University of Sydney, NSW 2006, Australia}

\begin{abstract}
Spray atomization process involves complex multi-phase phenomena. Abundant literature and validation of spray modelling for industrial applications like fuel injection in internal combustion and turbine jet engines are available. However, only a handful of studies, primarily limited to discrete phase modelling, of low-pressure applications, such as nasal spray exists. This study aims to provide insight into the external and near-nozzle spray characterization of a continuous spray and establishes good validation against the experiment.

A 3-dimensional (3D) X-ray scanner was used to extract the internal nasal spray nozzle geometry which was reconstructed to build a 3D computational model. A novel volume-of-fluid to discrete phase transition model (VOF-DPM) was used to track the liquid phase and its transition to droplets, which was based on the shape and size of the liquid lumps.

In this study, an early pre-stable and stable phase of spray plume development was investigated. Qualitative and quantitative analysis were carried out to validate the computational model. A liquid column exited a nozzle which distorted at its base with advancement in time and eventually formed a hollow-cone liquid sheet. It then, disintegrated due to instability that produced fluctuations to form ligaments resulting in secondary break-up.

This study provides in-depth understanding of liquid jet disintegration and droplet formation, which adds value to future nasal spray device designs and techniques to facilitate more effective targeted nasal drug delivery.

\end{abstract}

\begin{keyword}
nasal cavity \sep nasal spray \sep primary break-up \sep secondary break-up \sep computational fluid dynamics \sep VOF-DPM transition \sep CFD \sep nasal deposition

\end{keyword}

\end{frontmatter}

%% file: main.bbl
\begin{thebibliography}{53}
\expandafter\ifx\csname natexlab\endcsname\relax\def\natexlab#1{#1}\fi
\providecommand{\url}[1]{\texttt{#1}}
\providecommand{\href}[2]{#2}
\providecommand{\path}[1]{#1}
\providecommand{\DOIprefix}{doi:}
\providecommand{\ArXivprefix}{arXiv:}
\providecommand{\URLprefix}{URL: }
\providecommand{\Pubmedprefix}{pmid:}
\providecommand{\doi}[1]{\href{http://dx.doi.org/#1}{\path{#1}}}
\providecommand{\Pubmed}[1]{\href{pmid:#1}{\path{#1}}}
\providecommand{\bibinfo}[2]{#2}
\ifx\xfnm\relax \def\xfnm[#1]{\unskip,\space#1}\fi
\bibitem[{Amini(2016)}]{Amini2016}
\bibinfo{author}{Amini, G.}, \bibinfo{year}{2016}.
\newblock \bibinfo{title}{Liquid flow in a simplex swirl nozzle}.
\newblock \bibinfo{journal}{International Journal of Multiphase Flow}
  \bibinfo{volume}{79}, \bibinfo{pages}{225--235}.
\bibitem[{Basu et~al.(2020)Basu, Holbrook, Kudlaty, Fasanmade, Wu, Burke,
  Langworthy, Farzal, Mamdani, Bennett et~al.}]{Basu2020}
\bibinfo{author}{Basu, S.}, \bibinfo{author}{Holbrook, L.T.},
  \bibinfo{author}{Kudlaty, K.}, \bibinfo{author}{Fasanmade, O.},
  \bibinfo{author}{Wu, J.}, \bibinfo{author}{Burke, A.},
  \bibinfo{author}{Langworthy, B.W.}, \bibinfo{author}{Farzal, Z.},
  \bibinfo{author}{Mamdani, M.}, \bibinfo{author}{Bennett, W.D.}, et~al.,
  \bibinfo{year}{2020}.
\newblock \bibinfo{title}{Numerical evaluation of spray position for improved
  nasal drug delivery}.
\newblock \bibinfo{journal}{Scientific reports} \bibinfo{volume}{10},
  \bibinfo{pages}{1--18}.
\bibitem[{Brackbill et~al.(1992)Brackbill, Kothe and Zemach}]{Brackbill1992}
\bibinfo{author}{Brackbill, J.U.}, \bibinfo{author}{Kothe, D.B.},
  \bibinfo{author}{Zemach, C.}, \bibinfo{year}{1992}.
\newblock \bibinfo{title}{A continuum method for modeling surface tension}.
\newblock \bibinfo{journal}{Journal of computational physics}
  \bibinfo{volume}{100}, \bibinfo{pages}{335--354}.
\bibitem[{Calmet et~al.(2019)Calmet, Inthavong, Eguzkitza, Lehmkuhl, Houzeaux
  and V{\'a}zquez}]{Calmet2019}
\bibinfo{author}{Calmet, H.}, \bibinfo{author}{Inthavong, K.},
  \bibinfo{author}{Eguzkitza, B.}, \bibinfo{author}{Lehmkuhl, O.},
  \bibinfo{author}{Houzeaux, G.}, \bibinfo{author}{V{\'a}zquez, M.},
  \bibinfo{year}{2019}.
\newblock \bibinfo{title}{Nasal sprayed particle deposition in a human nasal
  cavity under different inhalation conditions}.
\newblock \bibinfo{journal}{PloS one} \bibinfo{volume}{14},
  \bibinfo{pages}{e0221330}.
\bibitem[{Charalampous et~al.(2019)Charalampous, Hadjiyiannis and
  Hardalupas}]{charalampous2019proper}
\bibinfo{author}{Charalampous, G.}, \bibinfo{author}{Hadjiyiannis, C.},
  \bibinfo{author}{Hardalupas, Y.}, \bibinfo{year}{2019}.
\newblock \bibinfo{title}{Proper orthogonal decomposition of primary breakup
  and spray in co-axial airblast atomizers}.
\newblock \bibinfo{journal}{Physics of Fluids} \bibinfo{volume}{31},
  \bibinfo{pages}{043304}.
\bibitem[{Cheng et~al.(2001)Cheng, Holmes, Gao, Guilmette, Li, Surakitbanharn
  and Rowlings}]{Cheng2001}
\bibinfo{author}{Cheng, Y.}, \bibinfo{author}{Holmes, T.},
  \bibinfo{author}{Gao, J.}, \bibinfo{author}{Guilmette, R.},
  \bibinfo{author}{Li, S.}, \bibinfo{author}{Surakitbanharn, Y.},
  \bibinfo{author}{Rowlings, C.}, \bibinfo{year}{2001}.
\newblock \bibinfo{title}{Characterization of nasal spray pumps and deposition
  pattern in a replica of the human nasal airway}.
\newblock \bibinfo{journal}{Journal of Aerosol Medicine} \bibinfo{volume}{14},
  \bibinfo{pages}{267--280}.
\bibitem[{Chinn(1996)}]{Chinn1996}
\bibinfo{author}{Chinn, J.J.}, \bibinfo{year}{1996}.
\newblock \bibinfo{title}{The internal flow physics of swirl atomizer nozzles}.
\newblock Ph.D. thesis. University of Manchester: UMIST.
\bibitem[{Dayal et~al.(2004)Dayal, Shaik and Singh}]{Dayal2004}
\bibinfo{author}{Dayal, P.}, \bibinfo{author}{Shaik, M.S.},
  \bibinfo{author}{Singh, M.}, \bibinfo{year}{2004}.
\newblock \bibinfo{title}{Evaluation of different parameters that affect
  droplet-size distribution from nasal sprays using the malvern
  spraytec{\textregistered}}.
\newblock \bibinfo{journal}{Journal of pharmaceutical sciences}
  \bibinfo{volume}{93}, \bibinfo{pages}{1725--1742}.
\bibitem[{Di~Martino et~al.(2022)Di~Martino, Ahirwal and
  Maffettone}]{di2022computational}
\bibinfo{author}{Di~Martino, M.}, \bibinfo{author}{Ahirwal, D.},
  \bibinfo{author}{Maffettone, P.L.}, \bibinfo{year}{2022}.
\newblock \bibinfo{title}{Computational fluid dynamics characterization of the
  hollow-cone atomization: Newtonian and non-newtonian spray comparison}.
\newblock \bibinfo{journal}{Physics of Fluids} \bibinfo{volume}{34},
  \bibinfo{pages}{093318}.
\bibitem[{Dombrowski and Hasson(1969)}]{Dombrowski1969}
\bibinfo{author}{Dombrowski, N.}, \bibinfo{author}{Hasson, D.},
  \bibinfo{year}{1969}.
\newblock \bibinfo{title}{The flow characteristics of swirl (centrifugal) spray
  pressure nozzles with low viscosity liquids}.
\newblock \bibinfo{journal}{AIChE Journal} \bibinfo{volume}{15},
  \bibinfo{pages}{604--611}.
\bibitem[{Foo et~al.(2007)Foo, Cheng, Su and Donovan}]{Foo2007}
\bibinfo{author}{Foo, M.Y.}, \bibinfo{author}{Cheng, Y.S.},
  \bibinfo{author}{Su, W.C.}, \bibinfo{author}{Donovan, M.D.},
  \bibinfo{year}{2007}.
\newblock \bibinfo{title}{The influence of spray properties on intranasal
  deposition}.
\newblock \bibinfo{journal}{Journal of Aerosol Medicine} \bibinfo{volume}{20},
  \bibinfo{pages}{495--508}.
\bibitem[{Fraser(1953)}]{Fraser1953}
\bibinfo{author}{Fraser, R.}, \bibinfo{year}{1953}.
\newblock \bibinfo{title}{Research into the performance of atomizers for
  liquids}.
\newblock \bibinfo{journal}{Imp. Coll. Chem. Eng. Soc. J.} \bibinfo{volume}{7},
  \bibinfo{pages}{52--68}.
\bibitem[{Fung et~al.(2013)Fung, Inthavong, Yang, Lappas and Tu}]{Fung2013}
\bibinfo{author}{Fung, M.C.}, \bibinfo{author}{Inthavong, K.},
  \bibinfo{author}{Yang, W.}, \bibinfo{author}{Lappas, P.},
  \bibinfo{author}{Tu, J.}, \bibinfo{year}{2013}.
\newblock \bibinfo{title}{External characteristics of unsteady spray
  atomization from a nasal spray device}.
\newblock \bibinfo{journal}{Journal of Pharmaceutical Sciences}
  \bibinfo{volume}{102}, \bibinfo{pages}{1024--1035}.
\bibitem[{Gao et~al.(2020)Gao, Shen and Mao}]{Gao2020}
\bibinfo{author}{Gao, M.}, \bibinfo{author}{Shen, X.}, \bibinfo{author}{Mao,
  S.}, \bibinfo{year}{2020}.
\newblock \bibinfo{title}{Factors influencing drug deposition in the nasal
  cavity upon delivery via nasal sprays}.
\newblock \bibinfo{journal}{Journal of Pharmaceutical Investigation}
  \bibinfo{volume}{50}, \bibinfo{pages}{251--259}.
\bibitem[{Guo and Doub(2006)}]{Guo2006}
\bibinfo{author}{Guo, C.}, \bibinfo{author}{Doub, W.H.}, \bibinfo{year}{2006}.
\newblock \bibinfo{title}{The influence of actuation parameters on in vitro
  testing of nasal spray products}.
\newblock \bibinfo{journal}{Journal of pharmaceutical sciences}
  \bibinfo{volume}{95}, \bibinfo{pages}{2029--2040}.
\bibitem[{Guo et~al.(2008)Guo, Stine, Kauffman and Doub}]{Guo2008}
\bibinfo{author}{Guo, C.}, \bibinfo{author}{Stine, K.J.},
  \bibinfo{author}{Kauffman, J.F.}, \bibinfo{author}{Doub, W.H.},
  \bibinfo{year}{2008}.
\newblock \bibinfo{title}{{Assessment of the influence factors on in vitro
  testing of nasal sprays using Box-Behnken experimental design}}.
\newblock \bibinfo{journal}{European Journal of Pharmaceutical Sciences}
  \bibinfo{volume}{35}, \bibinfo{pages}{417--426}.
\newblock \DOIprefix\doi{10.1016/j.ejps.2008.09.001}.
\bibitem[{Hirt and Nichols(1981)}]{Hirt1981}
\bibinfo{author}{Hirt, C.W.}, \bibinfo{author}{Nichols, B.D.},
  \bibinfo{year}{1981}.
\newblock \bibinfo{title}{Volume of fluid (\mbox{VOF}) method for the dynamics
  of free boundaries}.
\newblock \bibinfo{journal}{Journal of computational physics}
  \bibinfo{volume}{39}, \bibinfo{pages}{201--225}.
\bibitem[{Hossainpour and Binesh(2009)}]{hossainpour2009investigation}
\bibinfo{author}{Hossainpour, S.}, \bibinfo{author}{Binesh, A.},
  \bibinfo{year}{2009}.
\newblock \bibinfo{title}{Investigation of fuel spray atomization in a di
  heavy-duty diesel engine and comparison of various spray breakup models}.
\newblock \bibinfo{journal}{Fuel} \bibinfo{volume}{88},
  \bibinfo{pages}{799--805}.
\bibitem[{Inthavong et~al.(2014)Inthavong, Fung, Tong, Yang and
  Tu}]{Inthavong2014}
\bibinfo{author}{Inthavong, K.}, \bibinfo{author}{Fung, M.C.},
  \bibinfo{author}{Tong, X.}, \bibinfo{author}{Yang, W.}, \bibinfo{author}{Tu,
  J.}, \bibinfo{year}{2014}.
\newblock \bibinfo{title}{High resolution visualization and analysis of nasal
  spray drug delivery}.
\newblock \bibinfo{journal}{Pharmaceutical research} \bibinfo{volume}{31},
  \bibinfo{pages}{1930--1937}.
\bibitem[{Inthavong et~al.(2011)Inthavong, Ge, Se, Yang and Tu}]{Inthavong2011}
\bibinfo{author}{Inthavong, K.}, \bibinfo{author}{Ge, Q.}, \bibinfo{author}{Se,
  C.M.}, \bibinfo{author}{Yang, W.}, \bibinfo{author}{Tu, J.},
  \bibinfo{year}{2011}.
\newblock \bibinfo{title}{Simulation of sprayed particle deposition in a human
  nasal cavity including a nasal spray device}.
\newblock \bibinfo{journal}{Journal of Aerosol Science} \bibinfo{volume}{42},
  \bibinfo{pages}{100--113}.
\bibitem[{Inthavong et~al.(2008)Inthavong, Tian, Tu, Yang and
  Xue}]{Inthavong2008}
\bibinfo{author}{Inthavong, K.}, \bibinfo{author}{Tian, Z.F.},
  \bibinfo{author}{Tu, J.}, \bibinfo{author}{Yang, W.}, \bibinfo{author}{Xue,
  C.}, \bibinfo{year}{2008}.
\newblock \bibinfo{title}{Optimising nasal spray parameters for efficient drug
  delivery using computational fluid dynamics}.
\newblock \bibinfo{journal}{Computers in biology and medicine}
  \bibinfo{volume}{38}, \bibinfo{pages}{713--726}.
\bibitem[{Inthavong et~al.(2012)Inthavong, Yang, Fung and Tu}]{Inthavong2012}
\bibinfo{author}{Inthavong, K.}, \bibinfo{author}{Yang, W.},
  \bibinfo{author}{Fung, M.C.}, \bibinfo{author}{Tu, J.}, \bibinfo{year}{2012}.
\newblock \bibinfo{title}{External and near-nozzle spray characteristics of a
  continuous spray atomized from a nasal spray device}.
\newblock \bibinfo{journal}{Aerosol Science and Technology}
  \bibinfo{volume}{46}, \bibinfo{pages}{165--177}.
\bibitem[{Kelly et~al.(2004)Kelly, Asgharian, Kimbell and Wong}]{Kelly2004}
\bibinfo{author}{Kelly, J.T.}, \bibinfo{author}{Asgharian, B.},
  \bibinfo{author}{Kimbell, J.S.}, \bibinfo{author}{Wong, B.A.},
  \bibinfo{year}{2004}.
\newblock \bibinfo{title}{Particle deposition in human nasal airway replicas
  manufactured by different methods. part i: Inertial regime particles}.
\newblock \bibinfo{journal}{Aerosol science and technology}
  \bibinfo{volume}{38}, \bibinfo{pages}{1063--1071}.
\bibitem[{Kimbell et~al.(2007)Kimbell, Segal, Asgharian, Wong, Schroeter,
  Southall, Dickens, Brace and Miller}]{Kimbell2007}
\bibinfo{author}{Kimbell, J.S.}, \bibinfo{author}{Segal, R.A.},
  \bibinfo{author}{Asgharian, B.}, \bibinfo{author}{Wong, B.A.},
  \bibinfo{author}{Schroeter, J.D.}, \bibinfo{author}{Southall, J.P.},
  \bibinfo{author}{Dickens, C.J.}, \bibinfo{author}{Brace, G.},
  \bibinfo{author}{Miller, F.J.}, \bibinfo{year}{2007}.
\newblock \bibinfo{title}{Characterization of deposition from nasal spray
  devices using a computational fluid dynamics model of the human nasal
  passages}.
\newblock \bibinfo{journal}{Journal of aerosol medicine} \bibinfo{volume}{20},
  \bibinfo{pages}{59--74}.
\bibitem[{Kolanjiyil et~al.(2022)Kolanjiyil, Alfaifi, Aladwani, Golshahi and
  Longest}]{Kolanjiyil2022}
\bibinfo{author}{Kolanjiyil, A.V.}, \bibinfo{author}{Alfaifi, A.},
  \bibinfo{author}{Aladwani, G.}, \bibinfo{author}{Golshahi, L.},
  \bibinfo{author}{Longest, W.}, \bibinfo{year}{2022}.
\newblock \bibinfo{title}{Importance of spray--wall interaction and
  post-deposition liquid motion in the transport and delivery of pharmaceutical
  nasal sprays}.
\newblock \bibinfo{journal}{Pharmaceutics} \bibinfo{volume}{14},
  \bibinfo{pages}{956}.
\bibitem[{Kolanjiyil et~al.(2021)Kolanjiyil, Hosseini, Alfaifi, Hindle,
  Golshahi and Longest}]{Kolanjiyil2021}
\bibinfo{author}{Kolanjiyil, A.V.}, \bibinfo{author}{Hosseini, S.},
  \bibinfo{author}{Alfaifi, A.}, \bibinfo{author}{Hindle, M.},
  \bibinfo{author}{Golshahi, L.}, \bibinfo{author}{Longest, P.W.},
  \bibinfo{year}{2021}.
\newblock \bibinfo{title}{Importance of cloud motion and two-way momentum
  coupling in the transport of pharmaceutical nasal sprays}.
\newblock \bibinfo{journal}{Journal of Aerosol Science} \bibinfo{volume}{156},
  \bibinfo{pages}{105770}.
\bibitem[{Kumar and Sahu(2020)}]{kumar2020liquid}
\bibinfo{author}{Kumar, A.}, \bibinfo{author}{Sahu, S.}, \bibinfo{year}{2020}.
\newblock \bibinfo{title}{Liquid jet disintegration memory effect on downstream
  spray fluctuations in a coaxial twin-fluid injector}.
\newblock \bibinfo{journal}{Physics of Fluids} \bibinfo{volume}{32},
  \bibinfo{pages}{073302}.
\bibitem[{Kuo and Trujillo(2022)}]{kuo2022simulation}
\bibinfo{author}{Kuo, C.W.}, \bibinfo{author}{Trujillo, M.F.},
  \bibinfo{year}{2022}.
\newblock \bibinfo{title}{Simulation of liquid jet atomization and droplet
  breakup via a volume-of-fluid lagrangian--eulerian strategy}.
\newblock \bibinfo{journal}{Physics of Fluids} \bibinfo{volume}{34},
  \bibinfo{pages}{113326}.
\bibitem[{Lefebvre and McDonell(2017)}]{lefebvre2017atomization}
\bibinfo{author}{Lefebvre, A.H.}, \bibinfo{author}{McDonell, V.G.},
  \bibinfo{year}{2017}.
\newblock \bibinfo{title}{Atomization and sprays}.
\newblock \bibinfo{publisher}{CRC press}.
\bibitem[{Liu et~al.(1993)Liu, Mather and Reitz}]{Liu1993}
\bibinfo{author}{Liu, A.B.}, \bibinfo{author}{Mather, D.},
  \bibinfo{author}{Reitz, R.D.}, \bibinfo{year}{1993}.
\newblock \bibinfo{title}{Modeling the effects of drop drag and breakup on fuel
  sprays}.
\newblock \bibinfo{journal}{SAE Transactions} , \bibinfo{pages}{83--95}.
\bibitem[{Liu et~al.(2010)Liu, Doub and Guo}]{Liu2010}
\bibinfo{author}{Liu, X.}, \bibinfo{author}{Doub, W.H.}, \bibinfo{author}{Guo,
  C.}, \bibinfo{year}{2010}.
\newblock \bibinfo{title}{Evaluation of droplet velocity and size from nasal
  spray devices using phase doppler anemometry (\mbox{PDA})}.
\newblock \bibinfo{journal}{International journal of pharmaceutics}
  \bibinfo{volume}{388}, \bibinfo{pages}{82--87}.
\bibitem[{Liu et~al.(2017)Liu, Huang and Sun}]{Liu2017}
\bibinfo{author}{Liu, Z.}, \bibinfo{author}{Huang, Y.}, \bibinfo{author}{Sun,
  L.}, \bibinfo{year}{2017}.
\newblock \bibinfo{title}{Studies on air core size in a simplex pressure-swirl
  atomizer}.
\newblock \bibinfo{journal}{International Journal of Hydrogen Energy}
  \bibinfo{volume}{42}, \bibinfo{pages}{18649--18657}.
\bibitem[{Moon et~al.(2007)Moon, Bae, Abo-Serie and Choi}]{Moon2007}
\bibinfo{author}{Moon, S.}, \bibinfo{author}{Bae, C.},
  \bibinfo{author}{Abo-Serie, E.}, \bibinfo{author}{Choi, J.},
  \bibinfo{year}{2007}.
\newblock \bibinfo{title}{Internal and near-nozzle flow of a pressure-swirl
  atomizer under varied fuel temperature}.
\newblock \bibinfo{journal}{Atomization and Sprays} \bibinfo{volume}{17},
  \bibinfo{pages}{529--550}.
\bibitem[{Newman et~al.(1988)Newman, Moren and Clarke}]{Newman1988}
\bibinfo{author}{Newman, S.}, \bibinfo{author}{Moren, F.},
  \bibinfo{author}{Clarke, S.}, \bibinfo{year}{1988}.
\newblock \bibinfo{title}{Deposition pattern of nasal sprays in man.}
\newblock \bibinfo{journal}{Rhinology} \bibinfo{volume}{26},
  \bibinfo{pages}{111--120}.
\bibitem[{Patil and Sahu(2021)}]{patil2021air}
\bibinfo{author}{Patil, S.}, \bibinfo{author}{Sahu, S.}, \bibinfo{year}{2021}.
\newblock \bibinfo{title}{Air swirl effect on spray characteristics and droplet
  dispersion in a twin-jet crossflow airblast injector}.
\newblock \bibinfo{journal}{Physics of Fluids} \bibinfo{volume}{33},
  \bibinfo{pages}{073314}.
\bibitem[{Rajamanickam and Basu(2017)}]{Rajamanickam2017}
\bibinfo{author}{Rajamanickam, K.}, \bibinfo{author}{Basu, S.},
  \bibinfo{year}{2017}.
\newblock \bibinfo{title}{Insights into the dynamics of spray--swirl
  interactions}.
\newblock \bibinfo{journal}{Journal of Fluid Mechanics} \bibinfo{volume}{810},
  \bibinfo{pages}{82--126}.
\bibitem[{Rizk and Lefebvre(1987)}]{Rizk1987}
\bibinfo{author}{Rizk, N.}, \bibinfo{author}{Lefebvre, A.},
  \bibinfo{year}{1987}.
\newblock \bibinfo{title}{Prediction of velocity coefficient and spray cone
  angle for simplex swirl atomizers}.
\newblock \bibinfo{journal}{International Journal of Turbo and Jet Engines}
  \bibinfo{volume}{4}, \bibinfo{pages}{65--74}.
\bibitem[{Rizk and Lefebvre(1985)}]{Rizk1985}
\bibinfo{author}{Rizk, N.}, \bibinfo{author}{Lefebvre, A.H.},
  \bibinfo{year}{1985}.
\newblock \bibinfo{title}{Internal flow characteristics of simplex swirl
  atomizers}.
\newblock \bibinfo{journal}{Journal of propulsion and power}
  \bibinfo{volume}{1}, \bibinfo{pages}{193--199}.
\bibitem[{Rostami and Mahdavy~Moghaddam(2021)}]{rostami2021velocity}
\bibinfo{author}{Rostami, E.}, \bibinfo{author}{Mahdavy~Moghaddam, H.},
  \bibinfo{year}{2021}.
\newblock \bibinfo{title}{The velocity and viscosity impact on the annular
  spray atomisation of different fuels}.
\newblock \bibinfo{journal}{Combustion Theory and Modelling}
  \bibinfo{volume}{25}, \bibinfo{pages}{158--192}.
\bibitem[{Sahu et~al.(2022)Sahu, Chetan, Mahato, Kar, Das and
  Lakkaraju}]{sahu2022formation}
\bibinfo{author}{Sahu, T.L.}, \bibinfo{author}{Chetan, U.},
  \bibinfo{author}{Mahato, J.}, \bibinfo{author}{Kar, P.K.},
  \bibinfo{author}{Das, P.K.}, \bibinfo{author}{Lakkaraju, R.},
  \bibinfo{year}{2022}.
\newblock \bibinfo{title}{Formation and breakup of twisting ligaments in a
  viscous swirling liquid jet}.
\newblock \bibinfo{journal}{Physics of Fluids} \bibinfo{volume}{34},
  \bibinfo{pages}{112118}.
\bibitem[{Schmidt et~al.(1999)Schmidt, Nouar, Senecal, Rutland, Martin, Reitz
  and Hoffman}]{schmidt1999pressure}
\bibinfo{author}{Schmidt, D.P.}, \bibinfo{author}{Nouar, I.},
  \bibinfo{author}{Senecal, P.}, \bibinfo{author}{Rutland, J.},
  \bibinfo{author}{Martin, J.}, \bibinfo{author}{Reitz, R.D.},
  \bibinfo{author}{Hoffman, J.A.}, \bibinfo{year}{1999}.
\newblock \bibinfo{title}{Pressure-swirl atomization in the near field}.
\newblock \bibinfo{journal}{SAE transactions} , \bibinfo{pages}{471--484}.
\bibitem[{Shang et~al.(2021)Shang, Inthavong, Qiu, Singh, He and
  Tu}]{Shang2021}
\bibinfo{author}{Shang, Y.}, \bibinfo{author}{Inthavong, K.},
  \bibinfo{author}{Qiu, D.}, \bibinfo{author}{Singh, N.}, \bibinfo{author}{He,
  F.}, \bibinfo{author}{Tu, J.}, \bibinfo{year}{2021}.
\newblock \bibinfo{title}{Prediction of nasal spray drug absorption influenced
  by mucociliary clearance}.
\newblock \bibinfo{journal}{PLoS One} \bibinfo{volume}{16},
  \bibinfo{pages}{e0246007}.
\bibitem[{Shrestha et~al.(2020)Shrestha, Van~Strien, Singh and
  Inthavong}]{Shrestha2020}
\bibinfo{author}{Shrestha, K.}, \bibinfo{author}{Van~Strien, J.},
  \bibinfo{author}{Singh, N.}, \bibinfo{author}{Inthavong, K.},
  \bibinfo{year}{2020}.
\newblock \bibinfo{title}{Primary break-up and atomization characteristics of a
  nasal spray}.
\newblock \bibinfo{journal}{Plos One} \bibinfo{volume}{15},
  \bibinfo{pages}{e0236063}.
\bibitem[{Si et~al.(2021)Si, Sami and Xi}]{Si2021}
\bibinfo{author}{Si, X.A.}, \bibinfo{author}{Sami, M.}, \bibinfo{author}{Xi,
  J.}, \bibinfo{year}{2021}.
\newblock \bibinfo{title}{Liquid film translocation significantly enhances
  nasal spray delivery to olfactory region: a numerical simulation study}.
\newblock \bibinfo{journal}{Pharmaceutics} \bibinfo{volume}{13},
  \bibinfo{pages}{903}.
\bibitem[{Suman et~al.(2002)Suman, Laube, Lin, Brouet and Dalby}]{Suman2002}
\bibinfo{author}{Suman, J.D.}, \bibinfo{author}{Laube, B.L.},
  \bibinfo{author}{Lin, T.c.}, \bibinfo{author}{Brouet, G.},
  \bibinfo{author}{Dalby, R.}, \bibinfo{year}{2002}.
\newblock \bibinfo{title}{Validity of in vitro tests on aqueous spray pumps as
  surrogates for nasal deposition}.
\newblock \bibinfo{journal}{Pharmaceutical research} \bibinfo{volume}{19},
  \bibinfo{pages}{1--6}.
\bibitem[{Taylor(1948)}]{Taylor1948}
\bibinfo{author}{Taylor, G.I.}, \bibinfo{year}{1948}.
\newblock \bibinfo{title}{The mechanics of swirl atomizers}, in:
  \bibinfo{booktitle}{Seventh International congress of applied mechanics}, pp.
  \bibinfo{pages}{280--285}.
\bibitem[{Taylor(1963)}]{Taylor1963}
\bibinfo{author}{Taylor, G.I.}, \bibinfo{year}{1963}.
\newblock \bibinfo{title}{The shape and acceleration of a drop in a high speed
  air stream}.
\newblock \bibinfo{journal}{The scientific papers of GI Taylor}
  \bibinfo{volume}{3}, \bibinfo{pages}{457--464}.
\bibitem[{Tong et~al.(2016)Tong, Dong, Shang, Inthavong and Tu}]{Tong2016}
\bibinfo{author}{Tong, X.}, \bibinfo{author}{Dong, J.}, \bibinfo{author}{Shang,
  Y.}, \bibinfo{author}{Inthavong, K.}, \bibinfo{author}{Tu, J.},
  \bibinfo{year}{2016}.
\newblock \bibinfo{title}{Effects of nasal drug delivery device and its
  orientation on sprayed particle deposition in a realistic human nasal
  cavity}.
\newblock \bibinfo{journal}{Computers in biology and medicine}
  \bibinfo{volume}{77}, \bibinfo{pages}{40--48}.
\bibitem[{Treleaven et~al.(2022)Treleaven, Laera, Carmona, Odier, Gentil,
  Dombard, Daviller, Gicquel and Poinsot}]{treleaven2022coupling}
\bibinfo{author}{Treleaven, N.C.}, \bibinfo{author}{Laera, D.},
  \bibinfo{author}{Carmona, J.}, \bibinfo{author}{Odier, N.},
  \bibinfo{author}{Gentil, Y.}, \bibinfo{author}{Dombard, J.},
  \bibinfo{author}{Daviller, G.}, \bibinfo{author}{Gicquel, L.},
  \bibinfo{author}{Poinsot, T.}, \bibinfo{year}{2022}.
\newblock \bibinfo{title}{Coupling of combustion simulation with atomisation
  and filming models for les in swirled spray flames}.
\newblock \bibinfo{journal}{Flow, Turbulence and Combustion}
  \bibinfo{volume}{109}, \bibinfo{pages}{759--789}.
\bibitem[{Van~Strien et~al.(2022)Van~Strien, Petersen, Lappas, Yeo, Rezk,
  Vahaji and Inthavong}]{Van2022}
\bibinfo{author}{Van~Strien, J.}, \bibinfo{author}{Petersen, P.},
  \bibinfo{author}{Lappas, P.}, \bibinfo{author}{Yeo, L.},
  \bibinfo{author}{Rezk, A.}, \bibinfo{author}{Vahaji, S.},
  \bibinfo{author}{Inthavong, K.}, \bibinfo{year}{2022}.
\newblock \bibinfo{title}{Spray characteristics from nasal spray atomization}.
\newblock \bibinfo{journal}{Journal of Aerosol Science} \bibinfo{volume}{165},
  \bibinfo{pages}{106009}.
\bibitem[{York et~al.(1953)York, Stubbs and Tek}]{york1953mechanism}
\bibinfo{author}{York, J.L.}, \bibinfo{author}{Stubbs, H.},
  \bibinfo{author}{Tek, M.}, \bibinfo{year}{1953}.
\newblock \bibinfo{title}{The mechanism of disintegration of liquid sheets}.
\newblock \bibinfo{journal}{Transactions of the American society of Mechanical
  Engineers} \bibinfo{volume}{75}, \bibinfo{pages}{1279--1286}.
\bibitem[{Yu et~al.(2016)Yu, Li, Wang and Ding}]{yu2016modeling}
\bibinfo{author}{Yu, Y.}, \bibinfo{author}{Li, G.}, \bibinfo{author}{Wang, Y.},
  \bibinfo{author}{Ding, J.}, \bibinfo{year}{2016}.
\newblock \bibinfo{title}{Modeling the atomization of high-pressure fuel spray
  by using a new breakup model}.
\newblock \bibinfo{journal}{Applied Mathematical Modelling}
  \bibinfo{volume}{40}, \bibinfo{pages}{268--283}.
\bibitem[{Zheng et~al.(2023)Zheng, Zhao, Nian, Liu and
  Cheng}]{zheng2023numerical}
\bibinfo{author}{Zheng, L.}, \bibinfo{author}{Zhao, R.}, \bibinfo{author}{Nian,
  Y.L.}, \bibinfo{author}{Liu, J.}, \bibinfo{author}{Cheng, W.L.},
  \bibinfo{year}{2023}.
\newblock \bibinfo{title}{Numerical and experimental study of the effects of
  tangential to axial velocity ratio and structural parameters inside the
  nozzle on spray characteristics}.
\newblock \bibinfo{journal}{Physics of Fluids} \bibinfo{volume}{35},
  \bibinfo{pages}{043303}.

\end{thebibliography}
